# DeepVisInterests: CNN-Ontology Prediction of Users Interests from Social Images

Onsa Lazzez
Member of ReGIM-Lab
University of Sfax– Tunisia
onsa.lazzez.tn@ieee.org Wael Ouarda
Research Member of ReGIM- Lab
University of Sfax – Tunisia
wael.ouarda@ieee.org
Adel M. Alimi
Director of REGIM Lab
REsearch Groups –Professor
National Engineering School of Sfax– Tunisia
adel.alimi@ieee.org

Resum—In this paper, we present a novel system named DeepVisInterests that performs the users interests prediction task from social visual data based on a deep neural approach for the ontology construction. A comprehensive statistical study have been made to validate our DeepVisInterests system.
The proposed system is based on the construction of users interests ontology using a set of deep visual features in order to learn the semantic representation for the popular topics of interests defined by Facebook. In fact, DeepVisInterests system addressed the problem of discovering the attributed interests (how the user interest can be detected from her/his provided social images in OSN) and analyzing the performance of the automatic prediction by a comparison with the self-assessed topics of interests (topics of interests provided by user in a proposed questionnaire) through our experiments applied on social images database collected from 240 Facebook users.
The qualitative and the quantitative experimental study made in this paper, show that DeepVisInterests ranks top the list of recent related works with an accuracy of 0.80.

Index Terms—Online Social Networks, users interests, onto-logy, convolutional neural networks, object recognition.

## I. Introduction

The last decades have witnessed the boom of the online social networks (OSN) with the huge amount of social textual data (e.g. comments, tags, descriptions, etc.) and social visual data (e.g. liked images, shared images, etc.).
In fact, million of users named social users, visiting sites like Facebook, Pinterest, Instagram, Twitter, etc. These social networks sites principally rely on their social users to create and share data, called social data, in various types, to explain others' content with comments and to have a personal social graph based on on-line relationships. In fact, these social network sites continue to develop and social data continue to increase. In 2018, Facebook users have shared over 4 billion images with 5 billion textual data while Flickr users have shared over 3 billion images [35]. With the development of mobile device, the delivery of images become more convenient and the most popular social networks offer the chance to share images without limits with almost 300 million images shared daily on Facebook and around 60 million images on Instagram [10].

Yet, managing and understanding the social data is still an important research challenges [42] that are useful for diverse applications such as constructing better recommendation sys-tems and discovering social users' lifestyle, etc.
Thus, it has been well assumed that the social users are typically characterized by various features as personal attri-butes (demographic data, education, etc.), topic of interests, preferences, opinions, etc. For that, Kosinski et al. [29] and Lazzez et al. [23] confirm that the analysis of users' social data can be used to discover a set of their personal attributes such as age, gender, politic orientation, etc., using Facebook social data.
Furthermore, the social images understanding can be consi-dered as the image classification that allow to classify and to recover the class of an input image based on its objects. Therefore, an analysis of such shared images can be applied to generate an user' interest profile by analyzing the deep visual feature of their individual shared images and then aggregating the image-level information to predict user-level interest distribution at a fine-grained level. Evidently, as shown in figure 1, the users' interests discovery is either done in a static method, by collecting data that hardly changes, or in a dynamic method by collecting data that frequently changes. Actually, users' interests are presented explicitly by each user himself/ herself through likes and favourites (favourites books, movies, music, etc.) or implicitly by analyzing his/her social profile content.
Despite that, the most proposed works [29], [25] and [21] use the social textual data to analyze social networks and avoid the

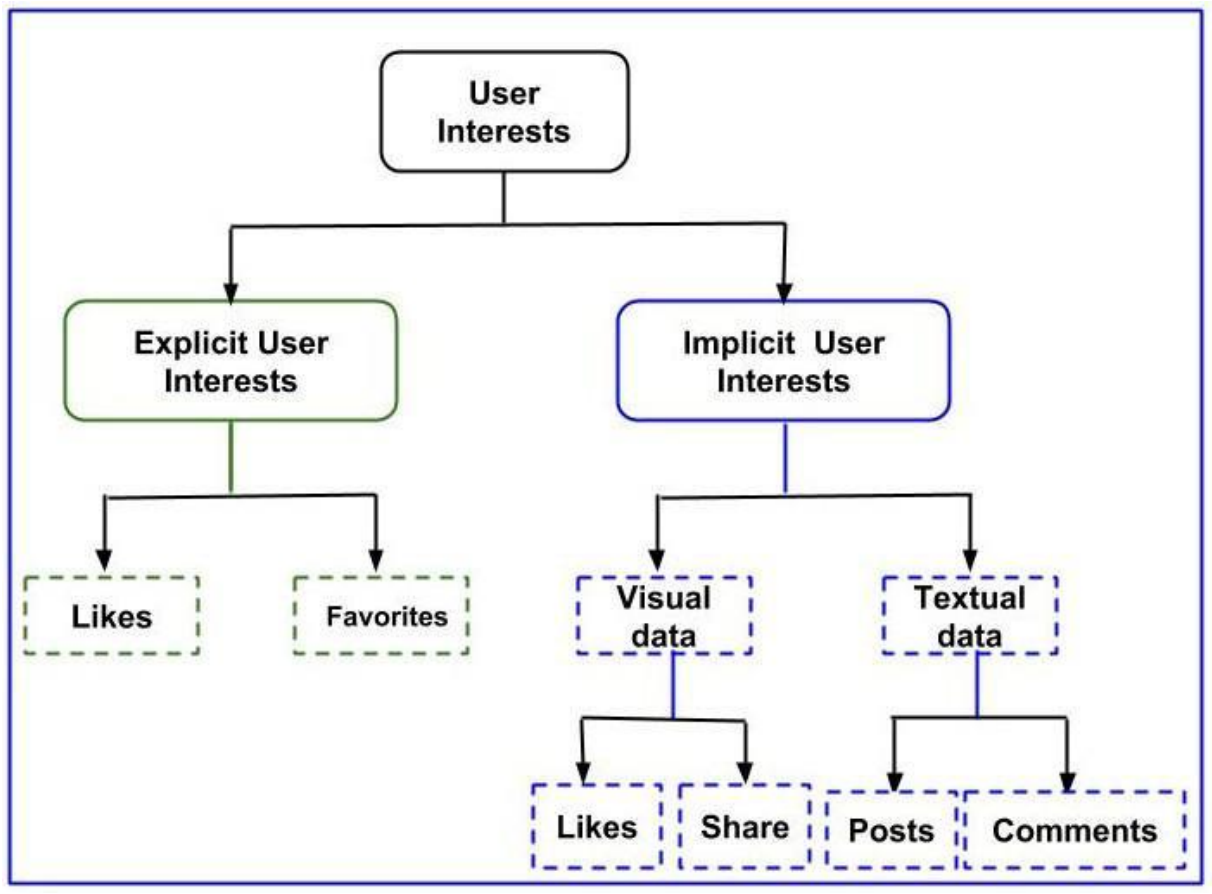

Fig. 1: Types of user interests

visual data, due to the limitation of available social images and the lack of social images benchmarks. Our proposed system is motivated by the key observation that images has become one of the most popular types of social data through which social users convey their personal attitudes likes preferences, personal data, topics of interests, etc.
However, users with diverse cultural environments, nationali-ties and ethnicities can easily communicate through a visual language to indicate their opinions, sentiments, personalities, etc [31], [48].
Due to the previous motivations, the social users' hidden information discovery process based on these data presents an attractive aspect [43] which is illustrated in the figure 2. The included objects on each image indicate the topic of interest such as places, family, culture which can indicate the topics in-terested by the user. Yet, the trends of users interests discovery from social visual data consists on the image understanding using object recognition technology. [3]. Indeed, the proposed object-based approach contains the following key steps: (a) the topics of interests classification and (b) the users' interest prediction. The reminder of this paper is structured as follow:

We constructed a novel social database named Smart-CityZenDB containing a set of social images gathered from 240 Facebook users accounts. This database was be annotated based on the self-assessed interests for each user in the set topics of interests defined by Facebook presented in table I.
For the users' interests conceptualization and categoriza-tion, we constructed a users' interests ontology based on the images' objects extracted from a set of benchmarks databases like FoodDB [33], SportDB [18], DeepFashi-onDB [27] and a set of our constructed database for the rest of mentioned topics. For the object recognition we used various CNN architectures like LeNet [24], AlexNet [22], GoogleNet [39] and VGG19 [36].
For the users' interest prediction, we infer the proposed ontology to result the attributed topics of interests for each user in the test database SmartCityZenDB. For that, various convolutional neural network architectures have been used to recognize the set of objects in each user' image that will be considered as sub-concepts in the ontology to predict their super-concepts.

This paper is structured as follow: The next section summa-rizes some papers from the literature about the techniques of social network analysis, of social visual data analysis using the CNN architectures, of users interests modelling and of users interests prediction basically from social visual data. Then, we will illustrate our proposed system DeepVisInterests giving details of each step including the object recognition architec-tures and the process of users interest ontology construction. The third section aims to illustrate experiments setup made in order to evaluate our DeepVisInterests framework. Finally, we will compare the set of CNN architectures for users interests discovery using the classification accuracy measure and we will discuss the found results on the new social database.

## II. State of the Art

There have been various related works on the analyzing of social networks from users' activities. This analysis can be used for many purposes, such as the recommendation systems, to discover hidden information about users in a fixed period of time, thus which can be useful information for psychologist to predict the personality of each user, the policymakers to classify trends in public opinions, etc.

### A. Social Network Analysis

This section explains the recent related works that apply social network analysis. The Social network analysis (SNA) is a concept that envelop descriptive and structure-based analysis. Victor Chang et al. [5] illustrate their social network analysis platform with their measurement methodology. They explain that social network analysis can perform the big data analytic. In fact, they describe the fundamental in their platform by using big data analytic tools: management, preprocessing and visualization to mine a large amount of social data from Facebook.
In [40], the authors proposed a novel system to apply the social network analysis tools for the egocentric Facebook network. They examined the relationships between a set of structural groups characteristics and individual perceptions of group cohesion. They have used a modularity algorithm and surveyed perceptions of cohesion and computed group density using visualization tools and Facebook data.
The authors in [12] proposed a novel system named HICODE to detect hidden communities from real world social network databases. Their system was based on some existing com-munity detection algorithms and other algorithms for finding disjoint communities and overlapping communities to reveal hidden structure. To validate their approach, they have used a set of real world dataset that are Facebook network like ( Caltech, Smith, Rice, Vassar, Wellesley,etc.) and SNAP network like (Youtube, Amazon and DBLP).

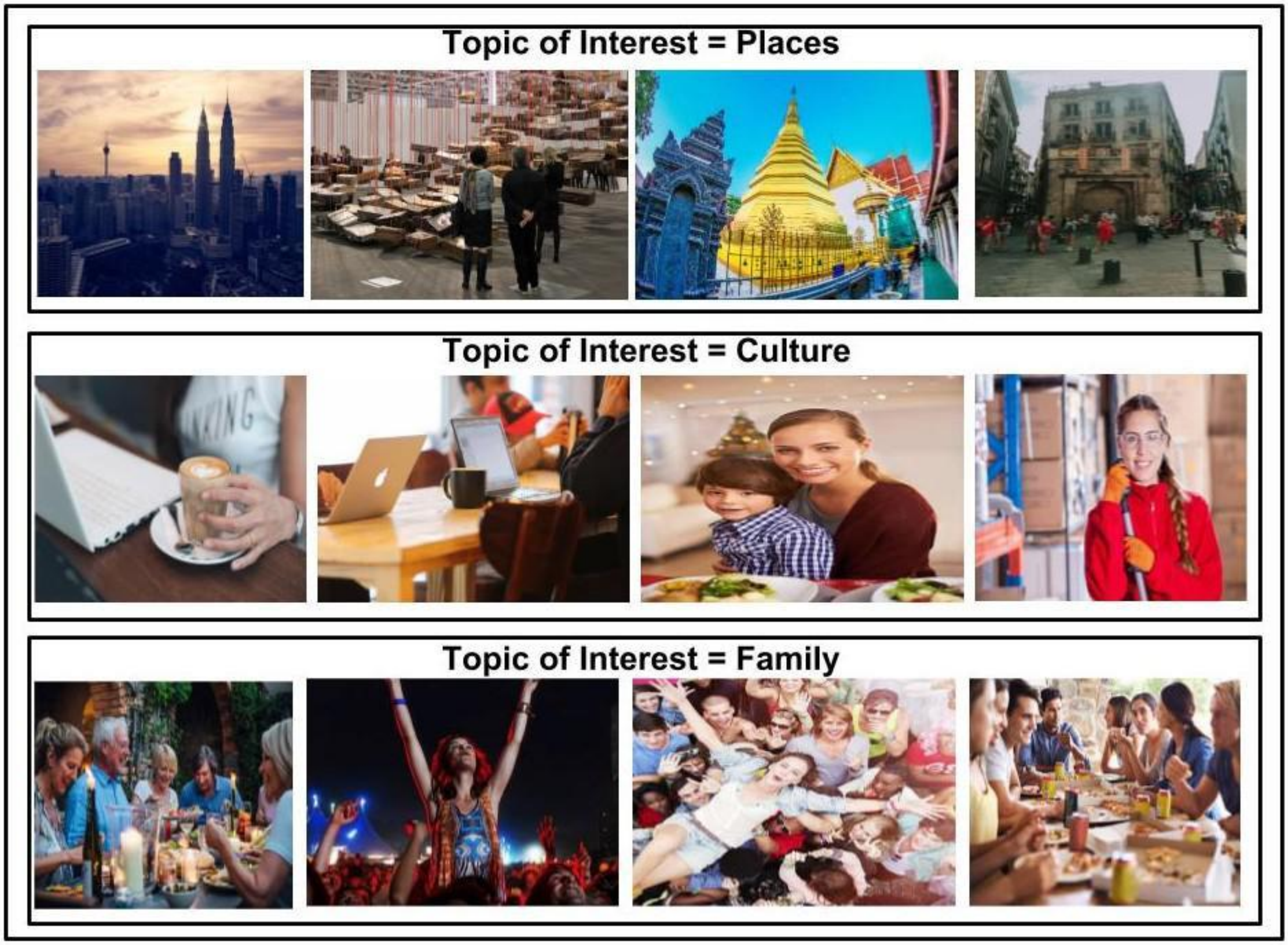

Fig. 2: Samples of Social images from one typical Facebook user

TABLE I: List of 24 core topics of Interests in Facebook

| | | | | | |
|---|---|---|---|---|---|
| Sport and Outdoors | Food and Drink | Shopping and Fashion | Fitness and Wellness | News and Entertainement | Business and Industry |
| Travel | Places and Events | Hobies and Activities | People | Technology | Family and Relationship |
| Education | Lifestyle and Culture | | | | |

B. Social Visual Data Analysis

Recent research approaches validates and highlight the idea of social information learning from social images [17]. Lazzez et al [23] proposed a novel system to discover demo-graphic attributes from social images. They applied the visual features and softmax classifier classify the users' demographic attributes such as age, gender, etc.from their profile pictures using a novel database from Facebook that contain 50 pictures.

Recently, with the important role of deep learning in image classification and the development of convolutional neural net-works (CNN) architectures, the social images content analysis has become much more adequate to map semantically the pixels to content of a given image [30].
Many architectures like AlexNet [22], LeNet [24],VGG19 [36] and GoogleNet [39], achieve the state of the art performance in the object recognition or detection applied on a huge amount of application areas.
Hence, the CNN as a feed-forward neural network, with various layers, tries to learns a hierarchical representation of the visual features from images.
Thus, based on CNN architectures, the authors in [34] propo-sed a novel level of image understanding by relating the visual features to personality traits. Their approach was focused on deep learning framework, in particular convolutional neural network (CNN) like AlexNet, VGG and Places. They applied these ImageNet pretrained CNN models and fine-tuning them to dicover personality traits by changing only the last layer in the network to adapt it to the multi-class classification problem. To validate their approach, they have employed a PsychoFlickr database that contains 300 Flick accounts and 200 random pictures for each user' account and they obtained an accuracy of 0,65.
Besides to personality analysis, the social visual data can be analyzed to predict users sentiments. IN [37], the aut-hors proposed a new system named SentiNet-A to integrate the visual attention into convolutional neural network deep sentiment classification framework. Their system was based on ImgeNet pretrained VGG to learn image representation, a multi layer neural network to predict the attention distribution of all image' regions toward image sentiment discovering and a multi-class to detect a general saliency map. In fact, the authors uses the saliency map of each image as regularize to refine each image' region attention distribtuion for sentiment prediction. To validate their approach, they have employed two databases that are Twitter database containing 1269 images and ArtPhotos databse that contains 806 artistic photos and they obtained an accuracy of 0,85 and 0,72 respectively for each mentioned database.

In the same context of sentiment analysis, the authors in [4] presented a novel methodology to move from pixels to sentiments. Their methodology was based on the transfer le-arning employing knowledge from pretrained network trained on ImageNet. In fact, they explained the fine-tuning AlexNet for visual sentiment analysis by fixing all the weights in the network except these in the last layers that are replaced by a new one with two neurones. To validate their methodology, they have used DeepSent database from twitter which contains 1269 images and they used the softmax classifier to obtain an accuray of 0,806.

### C. Users Interests Modeling

This section provides a brief overview of recent related works that extend the topic modeling approach for modeling user interests from social networks sites. In fact, the graph representation was the suitable tool for modeling semantic concepts from social data.

In [26], the authors proposed a Latent Topic of User Interests (LUI) model to model the topics distribution of tweets which possess non-Gaussian characteristics. They explained the full use of the non-Gaussian distribution of topics among tweets to uncover the users interests using a generative probabilistic approach. To evaluate their model, they have employed two microblogs Weibo and Twitter to construct two databases which contains 10 million tweets and 100 million tweets respectively and they obtained a correlation coefficient between topics of 0,64 and 0,63 for each mentioned database.

The authors in [44] presented a unified probabilistic user interests model named UIS-LDA. Tier model was based on Latent Dirichlet Allocation (LDA) as the most famous topic modeling algorithm using the library GLDA that utilize the GPU model in order to accelerate LDA on a single machine. To evaluate their model, the authors have used various database like Twitter database, Sina Weibo database and Epinion dataset and they obtained a F1 score of 0,0839, 0,1181 and 0,0365 respectively for each mentioned database.

The work proposed in [6] presented a user interests model based on Latent Dirichlet Allocation to model topics from forums to disting between the user's serious interest topics and the unserious interest topics. To validate their model, they have employed a forum threads from Tianya, a popular online forum site in China and they obtained an accuracies of 0,805 and 0,933 for serious ans unserious users.

### D. Users Interests Prediction

Some research efforts have been dedicated to improving the accuracy of social image topics of interests detection [15]. In [13] the authors proposed a method of assessing follow suggestions from social users based on categorical classification interests. They was based on the convolutional neural network architectures and a hierarchical topics of interests categorization. To validate their approach, they have used a database containing 1000 images collected from Instagram and they obtained a overall precision of 97,93%.

The work proposed by [46] presented a novel framework for user interests profiling from visual contents. The authors was based on Siamese network and convolutional neural network for featurtes extraction and the euclidean distance based soft assignment to pre-trained visual clusters in order to generate user profile by aggregating all image visual cluster features. To validate their framework, they have used 20,500 images collected from Pinterest and they have obtained a Mean Reciprocal Rank of 0.015 for 50 pins shared by each user.

In [28], Lovato et al. presented image classification based on content features obtained from deep learning method. This method was based on unsupervised learning algorithm learning image characteristics applied to online social networking ima-ge classification tasks. To verify the validity of the proposed method, they have used a novel database containing 5000 social images collected from Sina microblog and they obtained an accuracy of 89,70

In [49], the authors proposed a novel approach for the image and group level label propagation for users interests prediction based on the AlexNet architecture for deep visual features extraction and image level similarity to propagate the label information between images in order to propagate the ltopic of interest-evel knowledge for all user' images. To validate their approach, they have used a novel database collected from Pinterest containing 6000 images of 300 users accounts and they obtained an accuracy of 0,431%. In the same context, the table II illustrate the topics of interests used by some recent related works

## III. Proposed System

A social network is illustrated by a directed graph SoN = (H; F ) where the set H of nodes $h_i$ represents the set of social community H = $h_1$; $h_2$; :::; $h_H$ and the set F of links $f_{i;k}$ represents relationships between members as ordered pairs F = $f_{1;1}$; $f_{1;2}$; :::; $f_{jF\ j}$. On the other hand, the topics of interests are events ( posting, commenting, sharing, liking a post, publishing a post,etc.) illustrating a positive attention by a social user to a specific topic.

Because of the limited of social users energy, a social user has limited topics of interests. Among those interests, there are long-term and core interests and there must be temporary and marginal interests. Hence, Social users' core interests are stable and it will not change is a short period.

Social networks presents a lot of social visual data resulting complexity of the social users' activities. How to discover users' core interests from social visual data is the principal task of our work.

TABLE II: Topics of interests categories in the state of the art

| Work | Data Type | Topics |
|---|---|---|
| [20] | Tweets | Sport,Finance,Health,Movies and Digital |
| [19] | Tweets | Business, Politics, News, Home, Art, Society, Sports, Travel,Education, Fashion,Science, Careers, Religion, Technology, Food, Health,Pets, Family, Estate, Automotive, Finance, Shopping and Hobies |
| [49] | Images | Animals, Architecture, Art, Cars and Motorcycles, Celebrities, Design, Diy and Crafts, Education, Film and Movies and Book, Food and Drink, Gardering, Geek, Hair and Beauty, Health and Fitness, History, Holidays and Events, Home decor, Humor, Illustration and Posters, Kids, Men's fashion, Outdoors, Photography, Products, Quotes, Science and Nature, SPorts, Tattoos, Technology, Travel, Weddings, Women's fashion |
| [13] | Images | Arts, Business, \newline Computers,Games, Health, Home, News, Recreation,\newline Reference, Regional, Science, Shopping, Society, Sports, Kids,World |

However, within social network sites, various features make the prediction of user interests a crucial challenge. We focus on some challenges that influence the interest prediction process:

- Lack of social data in the explicit user profile: Each social user generally does not provide various information about his/her interest topic and the explicit user profile may never be considered fully known by a system. For that, we are not able to detect the user interests by analysing her/his profile directly.
- Social user behaviour: The social user is more and more active. Accordingly, detecting his/her topics of interests becomes difficult. In fact, the social user may define his/her interest about a specific topic through various data and then the choice of their type to analyze can be challenging.

Our approach aim to overcome these issues. For user activities and lack of information in the explicit user profile issues, the work focuses on the social user behaviour. The behaviour affect basically his/her shared images in order to benefit from a set of explicit deep visual features provided by the social user.

This paper invents a new deep framework presented in figure 3 that results each user' interests distribution among the core Facebook topics. This framework is based on an extensive study about extracting hidden information from social users' shared images. Particularly, to achieve high accuracy on this prediction, we applied the method of convolutional neural networks (CNN) to extract the deep visual features from social images. Currently, CNN has grained great success in image recognition and shows extensive performance improvement on classifying big image dataset such as ImageNet.[7].

## A. Problem Statement

Given a set of social users images as observations, we assume that each user is a mixture of a small number of core interests and each shared or likes image can be attributed to a specific topic.

Let U, I be the set of users and images for which the numbers are M and N respectively with $I_{N;m}$ is the $n^{th}$ image of user $U_m$.

Our goal is to predict the weight vector for each user that presents user-level-core-topics of interests-distribution among the 24 Facebook core topics of interests. In order to solve this challenge, we need to to understand the relationships between the topics using the users' interests modelling and detecting the images-level-user' interests-distribution by analyzing the social images' visual features based on the visual objects recognition.

## B. Convolutional Neural Network Architectures for Visual Objects Recognition

More lately, convolutional neural network architectures (CNN) have been appearing quickly. The great success of CNN in the ILSVRC 2017 challenges have clearly demonstrated the importance of these classifiers for extracting the characteristics contained within images [14]. Computer vision and pattern recognition have various breakthroughs due to the developement of CNN. CNN architectures provide an efficient tool to incorporate such primitive features into holistic representations, designed not only for highlight visual patterns but also to reveal information about the image content [45]. Basically, the CNN for image analysis posses two components: the convolution layers followed by the fully connected layers. The convolutional layers are what CNN characterize from the other type of neural networks. The objective of these layers is the extract the image' features. Behind the convolutional layers, the fully connected layers take over. The aim of these layers is to do classification. However, they classify the features that are extracted from the convolutional layers into various classes.

In this work, we applied models which are pre-trained on ImageNet Large Scale Visual Recignition Challenge (ILSVRC) dataset. The classification task of the last years' ILSVRC contains 1.2 million training images grouped into one thousand classes, which represent a wide variety of everyday objects.

1) ImageNet Database: ImageNet [8] is a large scale image database similar to the WordNet hierarchy. Each one of the hierarchy nodes is illustrated by hundreds and even thousands of images. Each important concept in WordNet, maybe are presented by various words or phrase and is named a "synset". Each synet is represented by 1000 images. In the following, we will give a simple summarize about the most known CNN architectures.

2) LeNet Architecture Details: First of all, LeNet [24], is a simple sequence of layers and every layer converts one volume

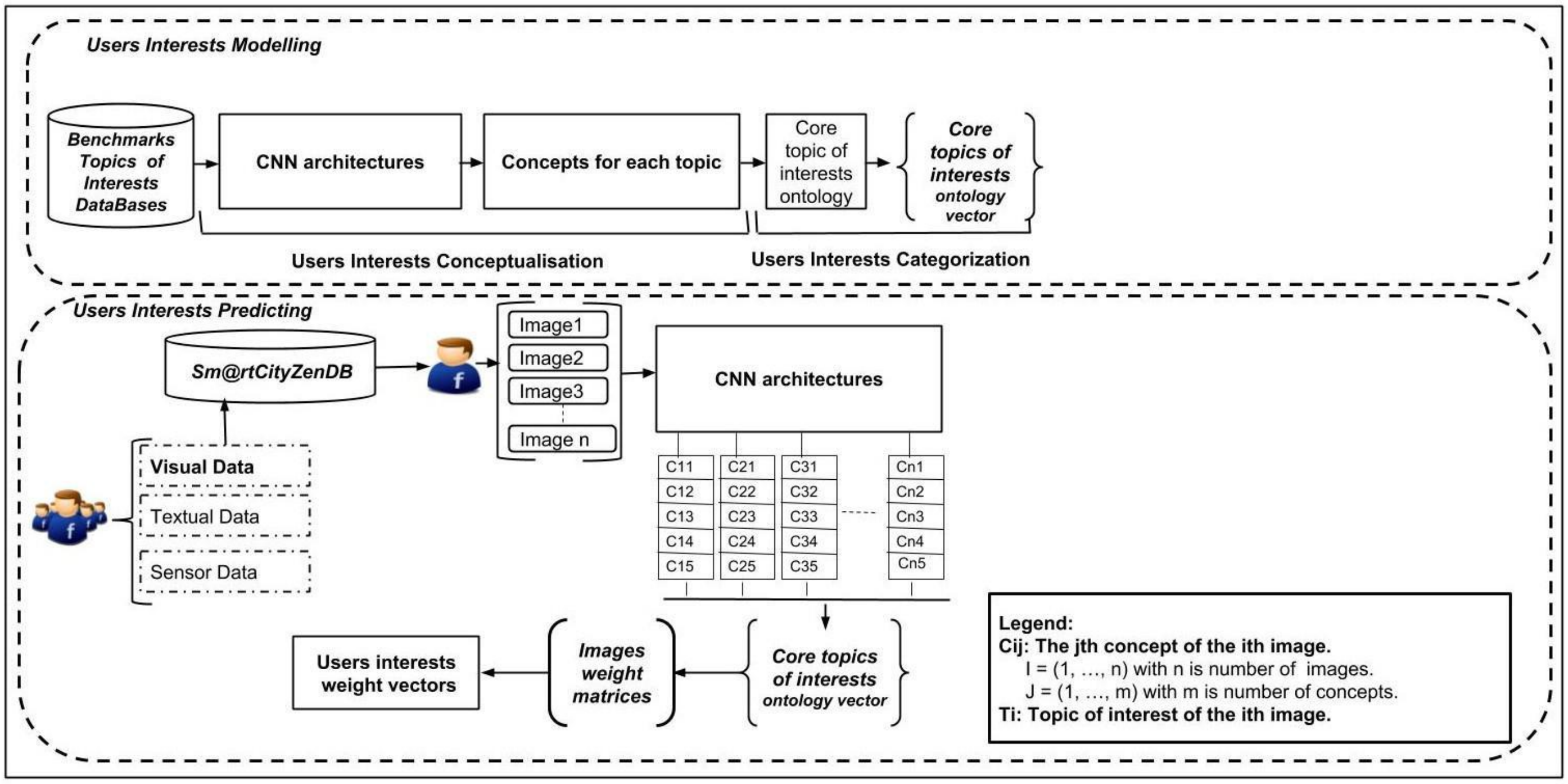

Fig. 3: DeepVisInterests: Deep Learning Framework for Users Interests discovery

of activation to another by a differentiable function. Let note that x is the input of each neuron in the LeNet architecture.

$$f(x) = x^{+} = \max(0; x) \quad (1)$$

The LeNet architecture is based on three principal types of layers that are Convolutional layer, Pooling layer and Fully-connected layer.

3) AlexNet Architecture Details: AlexNet [22], is the main architecture comes from the convolutional neural networks architectures, proposed by Krizhevsky et al., that has resulted the state of the art performance in the competition ImageNet classification challenge 2012. In fact, the layers in the AlexNet architecture are especially split into two blocks: the first block treat an input image in terms of visual features and the second block presents an important semantic representation based on those visual features [32]. The network had a very similar architecture to LeNet, but it was deeper, bigger and featured convolutional layers staked on top of each other. Before applying the ReLu function, AlexNet architecture use a response-normalized activity $b^{i}_{x;y}$ given by this expression:

$$b_{x;y}^{i} = a_{x;y}^{i} = (k + \sum_{j=\max(0;i\ n=2)}^{\min(N\ 1;i+n=2)} (a_{x;y}^{j})) \quad (2)$$

Actually, $a^{i}_{x;y}$ is the activity of a neuron computed by applying kernel i at position x; y and then applying the ReLu nonline-arity function.

4) VGG'19 Architecture Details: The VGG19 [36], is cha-racterized by its simplicity using only 3x3 convolutional layers stasked on top of each other in increasing depth. Reducing volume size is handled by max pooling. Two fully connected layers, each with 4,096 nodes are then followed by a softmax classifier.

5) GoogleNet Architecture Details: In addition to that, the GoogleNet [39] contains an average pooling layer with 5x5 filter size and stride 3, a 1x1 convolutional layer with 128 filters for dimension reduction and rectified linear activation, a fully connected layer with 1024 units and rectified linear activation.

Currently, table III illustrate our proposed comparison between the main CNN architectures based on the numbers of layers, filters and parameters for each architecture.

## C. Users Interests modelling

We present a novel semantic model, the user interests ontology (UIO) model, to capture the users interests from social visual data.

In fact, modelling users' interests present an important role in the current semantic web since it is at the basis of many services such as recommendation and customization. However, users' interests modelling help to reduce challenge such as domain-dependency, over-specialization and sparsity [9].

Our main technical contribution is presented from hierarc-hically users interests model using the ontology constructed by CNN for images classification. Indeed, ontology have been considered as an effective technique for modelling users interests.

, The ontology present the model related to a specific area [38]. It is used to organize data as a form of knowledge re-presentation. Individuals, called Instances, present the ground level of an ontology. Theses instances are then generalized into structures or Classes. A class in an ontology may be referred to as a type, concept or category, related to the ontology domain. In fact, each class can be subsumed by each other class. The subsming process define the class hierarchy and the concept of super-class and sub-class [16].

TABLE III: CNN Architectures comparison

| Architecture/ Features | Input Image | ConvLayer | Pooling Layer | FC | Size of kernels | Parameters |
|---|---|---|---|---|---|---|
| LeNet | 32*32*3 | 3 Convlayers | 2 average pooling | 2 FC | 6 (28*28) 6 (14*14) 6 (14*14) | 60M |
| AlexNet | 227*227*3 | 5 Convlayers | 3 max pooling | 3FC | 69 (11*11) 256 (5*5( 384 (3*3) 384 (3*3) 256 (3*3) | 60M |
| VGG19 | 224*224*3 | 16 Convlayers | 5 max pooling | 3 FC | (3*3) (2*2) | 140M |
| GoogleNet | 224*224*3 | 22 Convlayers | max pooling | No | 20 (1*1) (3*3) (5*5) | 4M |

Despite the important augmentation of visual data, many challenges remain in ranking computers capable to analyze visual features from images. Currently, ontology plays an efficient role in learning to classify set of images into general classes where there may not be a clear visual connect under class and image [11].

Actually, each image contains a set of various objects and each one may be used to classify it in the specific class. In order to convert the images objects to meaning, the principal challenge is to identify the pertinent concepts that both describe and identify each image. This challenge is resolved by a novel semi-automatic ontology based on deep learning techniques for images classification and web ontology language (OWL). In our ontology, the end nodes represents the 1000 classes of ImageNet, that we have obtained by fune-tuning CNN architectures on social images. Concerning the sub-nodes, we have generalized these 1000 classes to get the users interests in a hierarchical way.

The semi-automatic ontology construction is based on a set of benchmarks sources and other some our constructed databases, illustrated in figure 4, as the knowledge base to categorize the set of images' pertinent concepts under the core topics of interests. In this context, we propose two distinct problems: the extraction of the pertinent concepts from the images' objects and the construction of the ontology.

Relating to the first problem, we used the CNN models pretrained on ImageNet database. These models aim us to select the most pertinent images concepts from their 1000 objects such as plane, table, cat and other simple objects. For this purpose, we have selected AlexNet [22], LeNet [24], VGG [39] and GoogleNet [36] as the fundamental deep neural models. Furthermore, these models allow us to construct a deep learning based ontology for users' core interests con-ceptualization task presented on figure 5. For the second problem, after the object recognition, we use Protégé-OWL-5.2 to build the ontology, that is a free, open-source platform that provides tools to construct domain models and knowledge-based applications with ontology presented in figure 6.

However, CNN architectures are able to recognize simple fea-tures and objects from huge amount of images without giving any type of semantic meaning or creating any relationship between the various objects.

Figure 7 illustrate the incorporation of images' concepts, obtained by CNN architectures, as the input of our ontology to categorize each topic of interest. Each input image will be represented by a vector containing the set of scores of top 5 concepts among 1000 concepts of ImageNet[8]. The ontology vectorization method was used to found each user interest using OWL API. The vectorization method facilitates the semantic based negotiation and aims to make a vector that any of its elements represents a unique concept of ontology [37].

D. Metrics of Users Interests ontology

In this section, we present a set of metrics of our proposed ontology. An abstract model of users interests ontology is formally defined, using the structural similarities shared by all ontologies, to provide the definitions of metrics.

1) Abstract model of Users Interests ontology: The users interests ontology O is a tuple (C, I, A, R).

I is a collection of finite sets indexed by C with I = $I^c$jc 2 C.

A is a collection of a set of attributes with $A = A^c$jc 2 C. Each ' 2 $A^c$ is an attribute of concept c. The value of each attribute ' for an instance 2 $I^c$ of concept c, is presented by '( ). '( ) is either a data value or type T or an instance of concept c.

R is a set of binary relations on the set of concepts. R = $r_1$; $r_2$; :::; $r_k$. For each r 2 R and r $CXC^0$ , (c; $c^0$ ) 2 i.

In fact, the ontology implementation features present the num-ber of concepts, data properties, objects properties, individuals and axioms [50].

Our proposed ontology apply all the implementation features and the object properties are used to define the relationships between individuals. Furthermore, some interesting observati-ons are highlighted in table IV.

2) Vocabulary of users interests ontology: This subsection defines the basic metrics for the sizes of users interests ontology on various aspects.

The size of our ontology is defined as follow:

$size_C(O)$ = jjCjj,
$size_I(O)$ = P jj$I^c$jj,
$size_A(O)$ = c2C jj$A^c$ jj,
$size_R(O)$ = r2R jjrjj,
$size(O) = size_c(O) + size_I(O) + size_A(O) + size_R(O)$
.

Let O be the Users Interests ontology:
$Size_c(O) = 33$, $Size_I(O) = 5$, $Size_A(O) = 443$, $Size_R(O) = 32$ and $Size(O) = 1555$.

3) Structure of users interests ontology: Structural metrics are the most immense examines metrics in the ontology presentation, exactly, cohesion metrics that measure the degree of relatedness between concepts. Among the cohesion metrics, we find the Relation-based structural complexity. In fact, for each r 2 R we have some few structural metrics such as the number of Root nodes with N $RN_r(O)$ = jj$Root_r(O)$jj, the number of Leaf nodes with N $LN_r(O)$ = jj$Leaf_r(O)$jj,

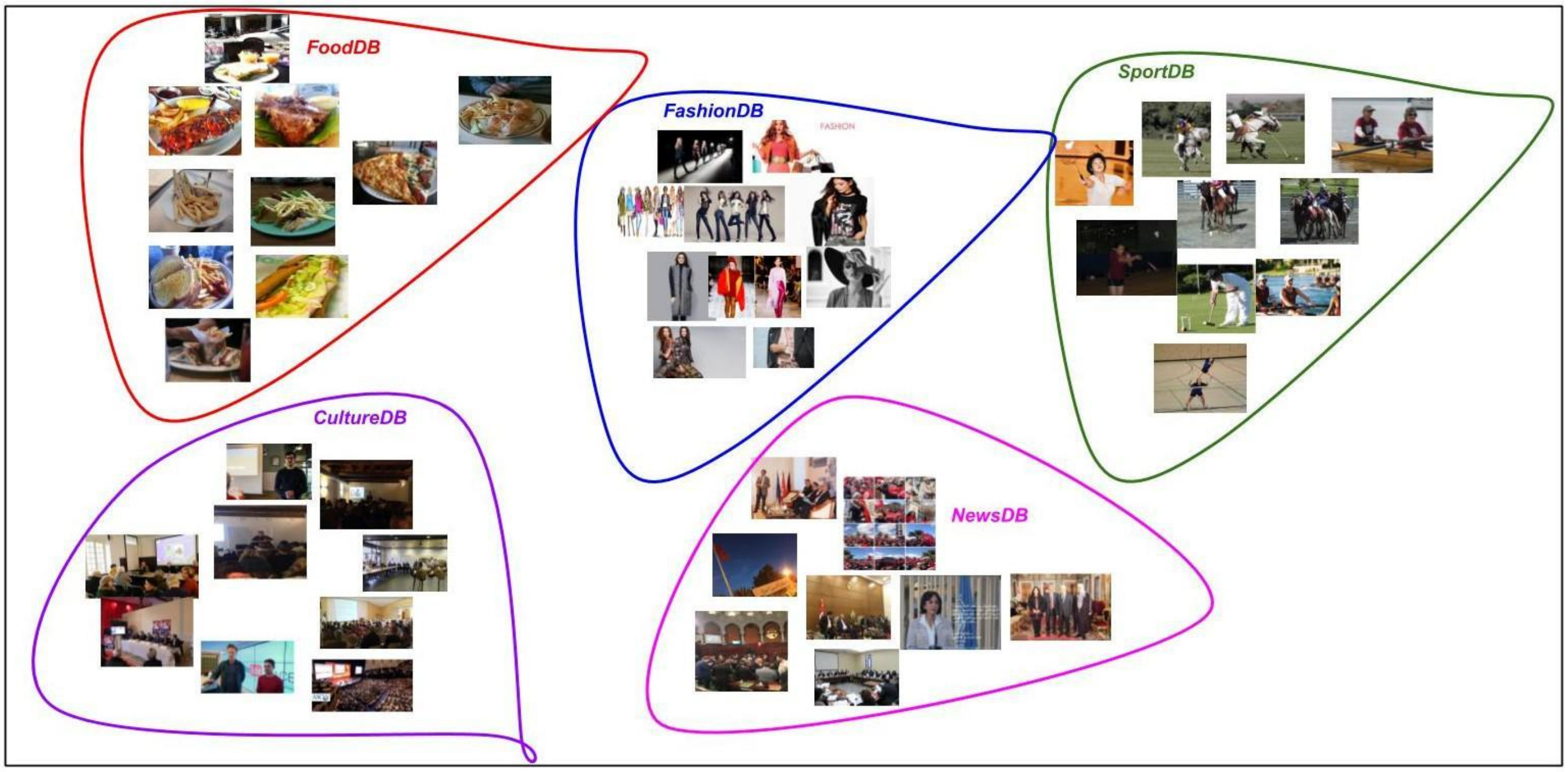

Fig. 4: Some External topics Databases

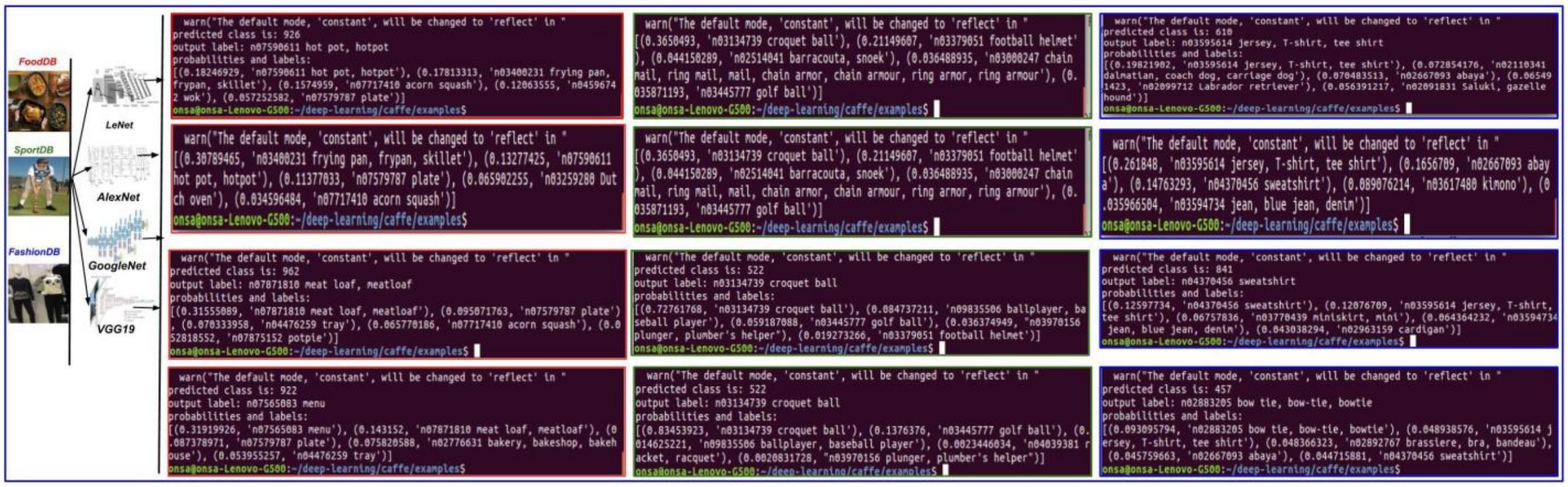

Fig. 5: Some core topics of interests conceptualization: Food, Sport and Fashion

TABLE IV: Users Interests Ontology Implementation Features

| Concepts | Object Proprieties | Data Proprieties | Individuals | Axioms |
|---|---|---|---|---|
| 33 | 32 | 0 | 443 | 1042 |

the maximum length of simple path with $MaxSPL_r(O) = Max_{p2path(O)}(Lenght(P))$,
the number of Isolated nodes as $NIC_r(O) = jjRoot_r(O) \; Leaf_r(O))jj$,
the total number of Reachable nodes from Roots with $TNRNR_r(O) = \; x2Root_r(O) \; jj \; Reachable_r(O) \; jj$ and
the average P

$ANRNR_r(O) = TNRNR_r(O) \; n \; jjNRN_r(O)jj$.

For the users interests ontology, the is-a relation based struc-ture metrics are:

$NRN = 1$, $NLN = 1$, $MaxSPL = 3$ and
$NIC = 0$, $TNRNT = 36$, $ANRNR = 36$.

4) Context of users interests ontology: We focus on users interests predicting and is interested with if an ontology is a perfect tool for modelling the semantics of topics of interests. Let Assume that a user U possesses n topics of interests $UI_1; :::; UI_n$ which contains a set of concepts $C_{i;1}; :::; C_{i;n}$ and a set of attributes $A_{i;1}; :::; A_{i;k}$.

interest consists of the following expressions:

$EXP_{UI_i}$ which describes the functionality of the topic $UI_i$,
$EXP_{C_{i;n}}$ which defines the meaning of the parameter $C_{i;j}$ in the set of concepts,
$EXP_{Y_{i;k}}$ which illustrates the meaning of the parameter

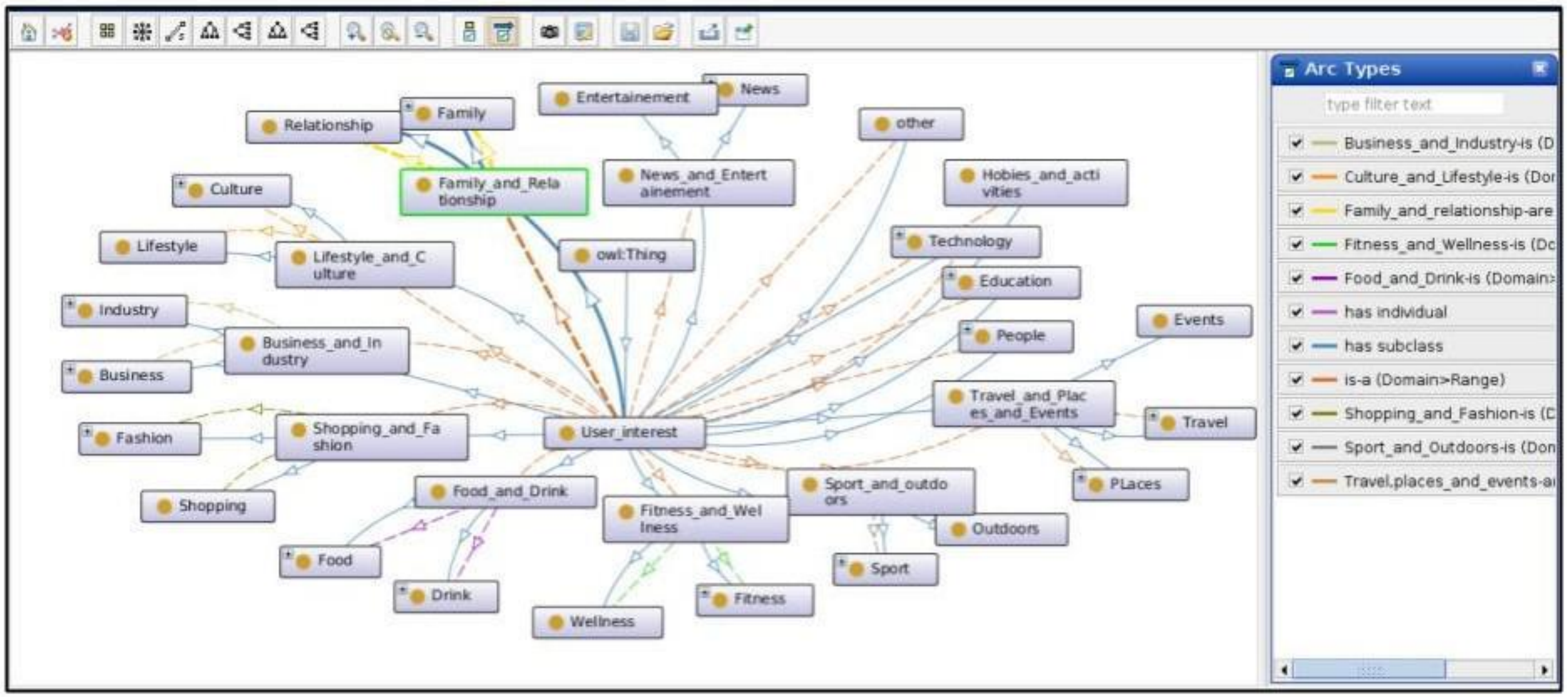

Fig. 6: Proposed Users Interests Ontology

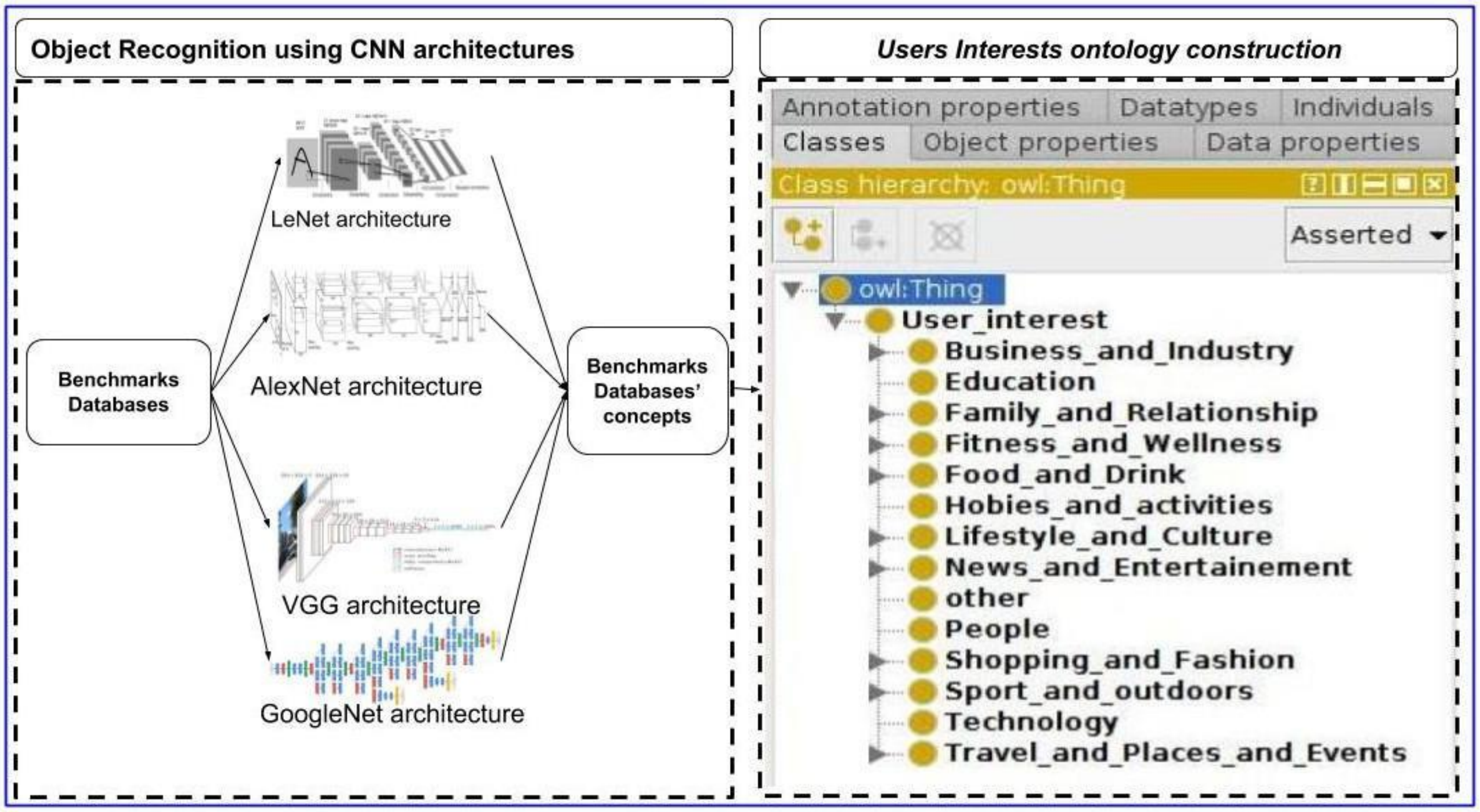

Fig. 7: Users Interests Ontology Design based CNN Architectures

$Y_{i;k}$, in the set of attributes of each concept.

5) Semiotic Metrics Assessment: The quality of each onto-logy is defined across a set of semiotic metrics. These metrics assess the syntactic, semantic, pragmatic and social aspects of ontology quality. The table V illustrate validation of these metrics on the users interests ontology.

## E. SmartCityZen Database: Pictures and Interests

Nowadays, creating database containing a big number of social images present an important challenge. In the social context, Facebook is the most popular social network that has been attracted attention from all over the world [1]. In fact, when a Facebook user creates a new account, a profile has to be constructed, which allows his family, friends and colleagues are able to identify himself. The profile include various types of social data. However, Facebook users provide divers types of data that can be very unsuitable, flexible and multiple such as personal data like (first name, last name, gender, location, education, family situation, etc.), preferences, favourites, likes, etc.

TABLE V: Semiotic Metrics Assessment Summary

| Aspect | Attributes | Description | Evaluation value |
|---|---|---|---|
| Syntactic | Lawfulness | Syntax/Correctness | 1 |
| | Richness | Syntax Breadth | 1 |
| Semantic | Interpretability | Terms Meaningfulness | 1 |
| | Consistency | Terms consistency | 1 |
| | Clarity | Terms clearness | 1 |
| Pragmatic | Comprehensiveness | Ontology size | 1550 |
| | Accuracy | Information Truthfulness | 1 |

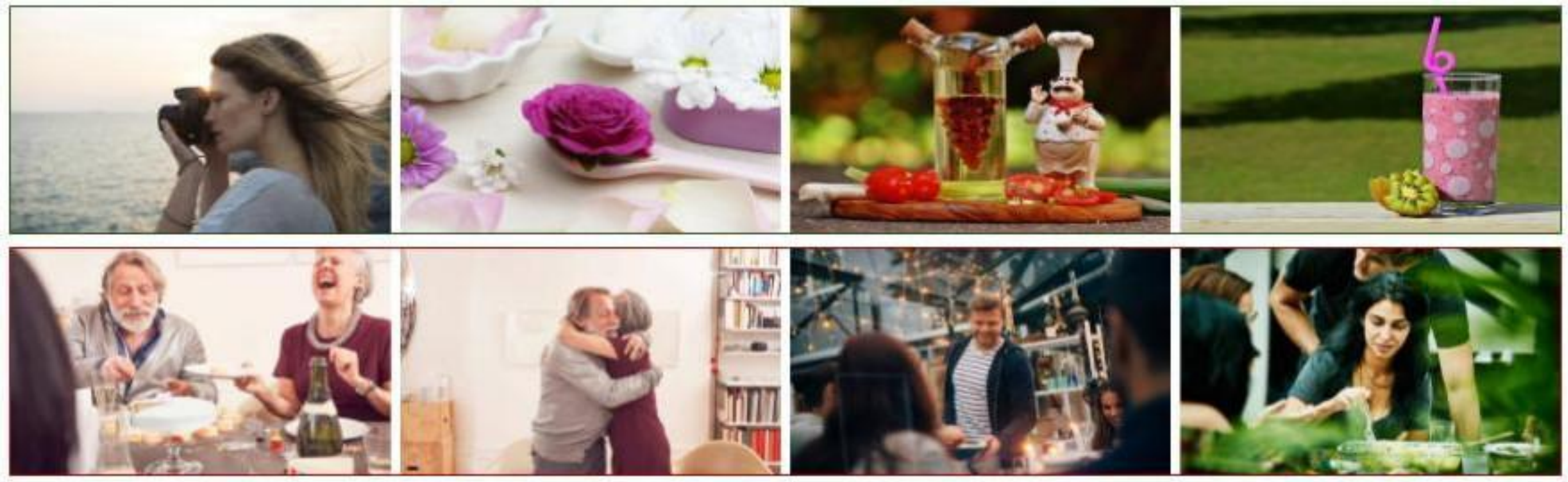

Fig. 8: Image samples from some topic of interests on SmartCityZenDB

However, collecting social data is not a trivial task as:

Facebook users can provide some data not publicly and Facebook users can not specify all personal data about themselves.

As social networks databases are not publicly accessible, the task of data collection from them attracted many researchers. [41].

Our main challenge is that social data collection from Facebook requires the implementation of a specific application compared to PInterest or Flicker that provides a public API. For this purpose, we developed a novel Facebook application named SmartCityZen platform [23], to collect Facebook users data. Each user have to connect with a currently Facebook account to give access to its data. In addition, each user can choose the data that can be collected by our platform such as: comments lists, liked pages, friends list, shared images, liked images, etc.

The sm@rtCityZen database contains social data on 240 Facebook accounts, simply called in the following social users. For each social user, there are 100 random images taken from those the "liked"and "shared". The database images have 320*320 resolution. It contains male and female users, mul-tiple ethnicity like Africans and European people in various locations. Their ages are within 15 and 60 years old. Figure 8 present some images samples from some topics of interests on Sm@rtCityZen database.

Furthermore, the database presents the self-assessed users interests. The former are based on a big interest questionnaire (BI) filled in voluntarily by each user. We note that the selected subsets are roughly balanced. The result of the BI is a vector where each component indicates the disposition of a user with respect to the core topics presented in table1. In fact, for each user U, we have a label $Self_{Up}$.
p = SandO; F andD; SandF; F andW; N andE; BandI; T andP andE; HandA; P; T; F andR; EandLandC, which is called the self-assessed core topics of interests signature. The self assessed traits are examined to be the validated user's core interests.

## F. Users Attributed Core Interests Prediction

Our approach is defined from the hypothesis that social data and mainly visual data product a various appropriate knowledge from which users interests can be extracted. After building the semantic users' interests model, we propose a visual users interests prediction based on the combination between image level and user level (VUIP-IL/UL) methods. Technically, this prediction comprises two main phases. Phase 1 is for the image level user interests distribution and phase 2 presents the user level interests distribution to obtain the target user interests matrix toward the Facebook core twenty-four distinct topics.

Figure 9 illustrates the main steps of our proposed method VUIP-IL/UL. In the first step, we apply the CNN architectures for objects recognition to extract the most deep visual features from each user' shared images from our test database.These features presents the image' objects with their probabilities. Thus, by inferring our user interest ontology (UIO) model,each image' object is replaced by her corresponded super concept. In the second step, the probability based and occurrence based scoring mechanisms are applied to result the image-level users interests distribution matrices. In the third step, we build a mapping matrix from the image-level distribution to the user-level distribution.

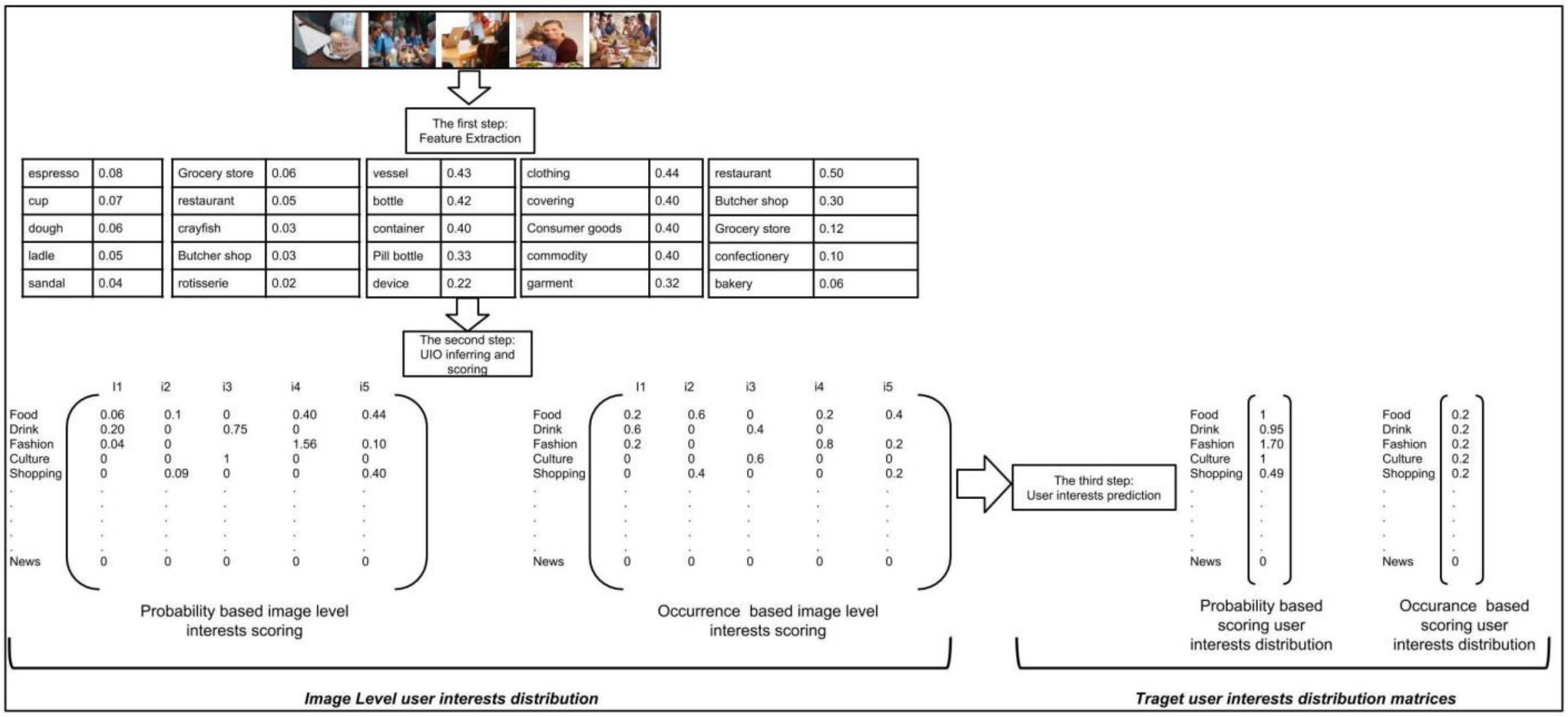

Fig. 9: An illustration of VUIP-IL/UL method. Three steps are incorporated.

Scoring Users Interests To quantify the user interests, a scoring function is proposed. The input of this scoring function examine a set of personal topics of interests that can be detected through social shared images. In fact, once the core user topics of interests are predicted, then we will score them to find the weight of each topics. The mechanism is very powerful as the user interests scores will be applied to determine the adapted interests distribution for each user' image thus for each user. We use a probability-based and occurrence based scoring mechanisms. The topic score of each user' image i 2 I, posted by a given user u 2 U, may be measured by probability and occurrence for an object o 2 O, where an image is represented by a collection of objects O.

$$S(u; I) = \sum_{i=1}^{n} p_{oi}\, oc_{oi} \qquad (3)$$

where $p_{oi}$ 2 $P_I$ , $oc_{oi}$ 2 $OC_I$ .

Here, $P_I$ is a set of probabilities within each i 2 I obtained by each image' objects and $OC_I$ is a set of occurrences with object o 2 O for the given image' user.

The algorithm 1 demonstrates the detailed steps used in our prediction task.

1) Image level interests distribution: Figure 10 presents the process of image level users interests distribution by illustrating the scoring mechanism.

After applying the feature extraction task, our image possesses 5 objects with their probabilities that are (espresso,0.08), (cup, 0.07), (dough, 0.06), (ladle, 0.05) and (sandal, 0.04). Using the Fact++ reasoner and DL query, we infer the user interests ontology to result the superclass for each image' object. We use the data property has-Instance in order to generate the superclass for each input image' object, presented as an ontology instance. This step apply the Fact++ reasoner and the DL query: (has-Instance value Ïmage' object"). For example (has-Instance value "espresso") result the super class Drink, (has-Instance value çup") result super class Drink, (has-Instance value "sandal") result superclass Fashion, (has-Instance value "dough") result super class Food and (has-Instance value ladle") result super class Drink.

For the scoring of topics for each image, we apply two mechanism presented in the figure 14. (a) For the probability based scoring, we define S(i; t) = P (o; i) where S(i; t) is the score of image i in the topic t with t 2 T and P (o; i) is the probability of object o of image i with o 2 O and O is the set of image' objects. (b) For the occurrence based scoring,

$S(i; t) = \frac{N(o;i;t)}{5}$ where N(o; i; t) is the number of image' objects with their super class in U IO are t.

In this way, we describe two matrices G $R^{n\ 24}$ and $G^0$ $R^{0n\ 24}$ to be the affinity matrices between the 24 core topics of interests and the n shared images by a specific user u U.

To give more details, Algorithm 2 explain the general steps for image-level-user-interests distribution.

2) User level interests distribution: According to the image level already explained, each user u 2 U possess two weighted matrices G and $G^0$ for n shared images in Facebook. For this level, we aim generate the target user interests distribution matrix based on the two scoring mechanisms which are (a) it first treats the matrix G, we define:

$$S(u; t_k) = \sum_{k=1}^{24}\sum_{i=1}^{n} p_{i;k} \qquad (4)$$

where $S(u; t_k)$ is the score of interest for the user u about topic t with $p_{i;k}$ is the probability of image i for topic k and n is the number of shared images.

---

Algorithm 1 User core Interests prediction

---

Require:

BD: List of 24 benchmarks databases presented the 24 Facebook core topics of interests.
Sm@rtCityZenDB: Test database.
P: Pretrained CNN models.
U: Users in test database.
I: collection of shared images of each u 2 U.

1) Extract the deep visual features from P of BD.
2) Users Interests Ontology construction: UIO.
3) UIO vectorisation.
4) Extract the deep visual features from P of I.
5) Image level interests distribution.
6) Return G, G': weight matrices specific for I of every u 2 U.
7) User level interests distribution.
8) Return V: weight vector specific for each u 2 U.

---

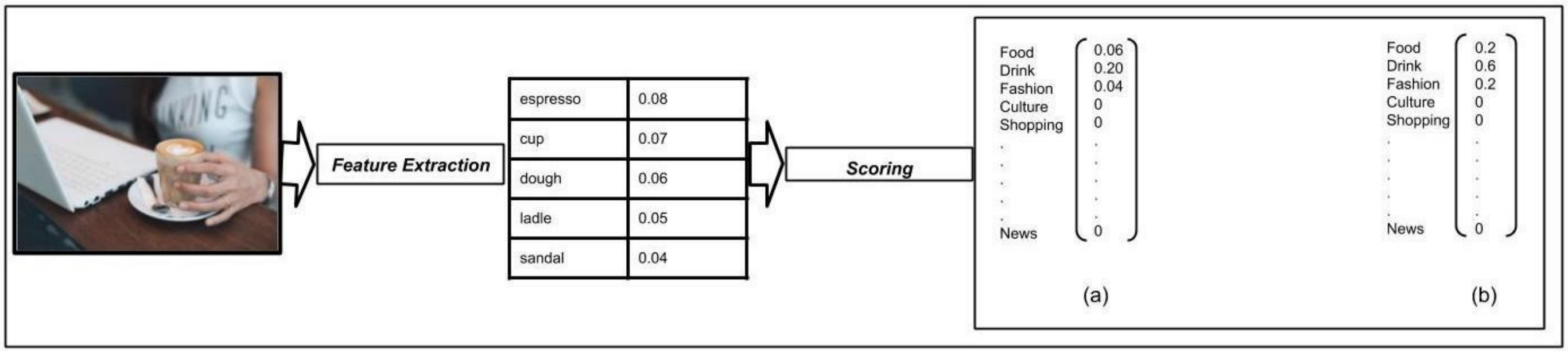

Fig. 10: An illustration of VUIP-IL method: Image Level Users Interests Distribution.

---

Algorithm 2 Image level users interests distribution

---

Require:

U: users in the test database.
UIO: users interests ontology.
T: List of Facebook 24 core topics of interests.
$I = i_1; i_2; :::; i_n$: n shared images by specific u U
P: Pretrained ImageNet CNN models.

1) Apply P for objects recognition on each i I to extract the five objects with high probability.
2) Infer UIO, using Fact++ reasoner and DL queries to predict the super class for every object per i I .
3) Scoring image-level-topics of interests.
4) Repeat :
5) Employ 1, 2 and 3 for all I.
6) Until n for the three cases: n = 5, n = 10 , n = 50, n = 75 and n = 100.
7) Return G, G': weight matrices of probabilities based and occurrence based scoring mechanism.

---

Algorithm 3 User level interests distribution

---

Require: U: users in the test database.
$I = i_1; i_2; :::; i_n$: n shared images by specific u U
G, G': weight matrices of probabilities based and occurrence based scoring mechanism respectively.

1) Extract target user interest distribution vector V for probability based scoring for the three cases: n = 5, n = 10 , n = 50, n = 75 and n = 100.
2) Extract target user interest distributions on vector V' for occurrence based scoring for the three cases: n = 5, n = 10 , n = 50, n = 75 and n = 100.
3) Return V,V'.

---

(b) the second mechanism treats the matrix $G^0$, we define for k=(1,24) and i=(1, n):

$$S(u; t_k) = \frac{N(\max(p_{ik}))}{n} \quad (5)$$

| Userid | Shared Images | Topics of Interests Distribution |
|---|---|---|
| U1 | 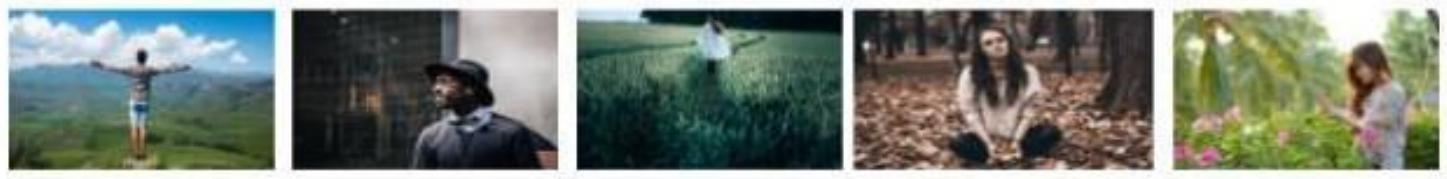 | Outdoors:0.8 PLaces :0.2 |
| U2 | 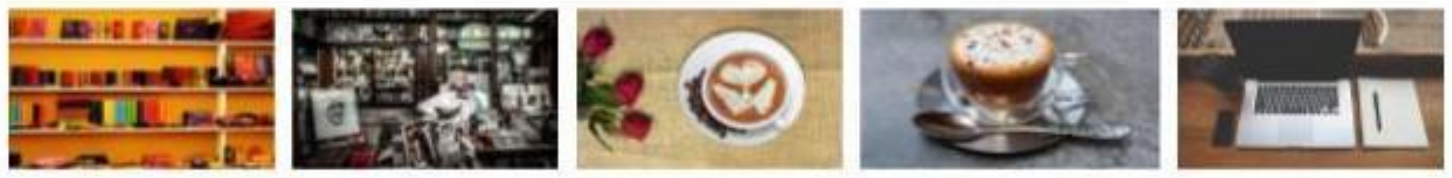 | Culture:0.6 Drink:0.4 |
| U3 | 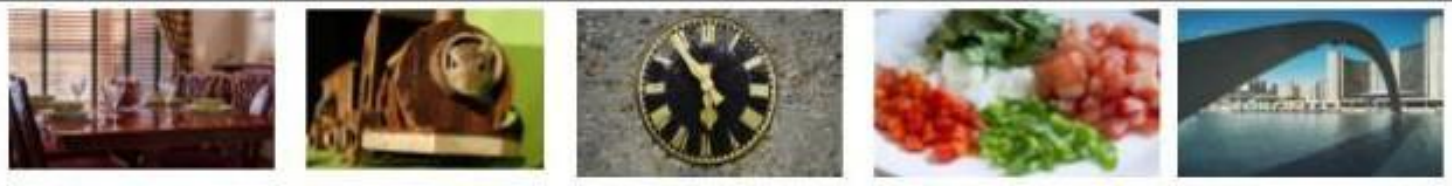 | Places:0.8 Food:0.2 |

Fig. 11: An illustration of VUIP-UL method: User Level Interests distribution for three selected users.

where $S(u; t_k)$ is the score of interest for the user u about topic t with $p_{i;k}$ is the probability of image i for topic k and n is the number of shared images. To give more details, algorithm 3 present the general steps of user interest distribution level. We choose three users u1,u2,u3 from our test database Sm@rtCityZenDB with 5 shared images as shown in figure 11.

In fact, Equation (6) shows the detail of aggregating the $i^{th}$ interest for user $U_m$ from his collection of images $I_{U;m}$.

$$I_{(U_m)_i} = \frac{\sum_{m \in I_{U_m}} P_m(i)}{\sum_k \sum_{m \in I_{U_m}} P_m(k)} : \quad (6)$$

## IV. Experimental Results

In this section, we first presents a qualitative study about the correlation between Facebook core topics of interests, we then illustrate a quantitative study about our obtained prediction results , we then demonstrate a comparison between the used pre-trained CNN models for object recognition and finally we highlight our framework DeepVisInterest by a comparison with the state of the art.

### A. Qualitative study on Facebook core topics of interests

Our proposed system DeepVisInterest facilitates the users' interests prediction based on the set of shared images by them. This prediction aims to define a degrees of correlation between some topics together. To calculate these degree, we use the Pearson Moment correlation to measure the dependency between various topics already mentioned in table1. We propose to learn the weighted matrix M from Facebook users' shared images. Specifically, the input of $M_{ij}$ is defined as the Pearson moment index between topics i and j. This index is the ratio between the users numbers who are interested by topics i and j in the same time and the users number who are interested by topic i or j. Figure 12 illustrate the coefficients between all different topics of interests. Through this figure, we can assume that Facebook users are interested by some topics together.

Definition : The Pearson Correlation Coefficient is the covariance of a population divided by the product of their standard deviation. It has a value between +1 and 1.

- where is between 1 and 0:5: is high total positive linear correlation,
- where is 0:5 and 0: is positive linear correlation,
- where is between 0 and 0:5: is negative linear correla-tion,
- and where is between 0:5 and 1 is high total negative linear correlation

This coefficient is commonly represented by p(rho) with:

$$p_{x;y} = \frac{Cov(X; Y)}{x; y} \quad (7)$$

where Cov is the covariance, $_x$ is the standard deviation of X and $_y$ is the standard deviation of Y .

In our work, p(rho) takes as input values the matrices from image-level interests distribution task and the figure 13, 14 and 15 illustrate the obtained correlation graphs.

Firstly, as illustrated in figure 13, we notice that the topic Food is high total positive correlated with the topics Drink, Family and People with a value grater than 0:5. This high correlation means that the user who is interested by the topic Food, is also interested by the Drink or Family or People and means that the images containing objects belonging the super class Food in our UIO ontology, may contain objects belonging the super class Drink or Family or People.

Secondly, as showing in figure 14, we observe that the topic Fashion is high negative correlated with topics Technology and Business with a value less than 0:5 which mean that the user who has Fashion as topic of interest can never be interested by topics Technology or Business. This high negativity is explained by the fact that each image corresponds to Fashion never contains objects belonging to super class Technology or Business in our UIO ontology.

Finally, through figure 15, the topic Education, for example, is correlated with topic Culture with a value between 0:5 and 1 means that a user interested by Culture is interested by Education and conversely for all topics which have a correlation value between 1 and 0:5.

### B. Quantitative study on user interests prediction

In our work we used our own annotated Facebook database with 240 users accounts distributed on 24 classes and each class contains 10 users. In this study we try to analyze the topics of interests prediction in the two levels : image-level and user-level already presented.

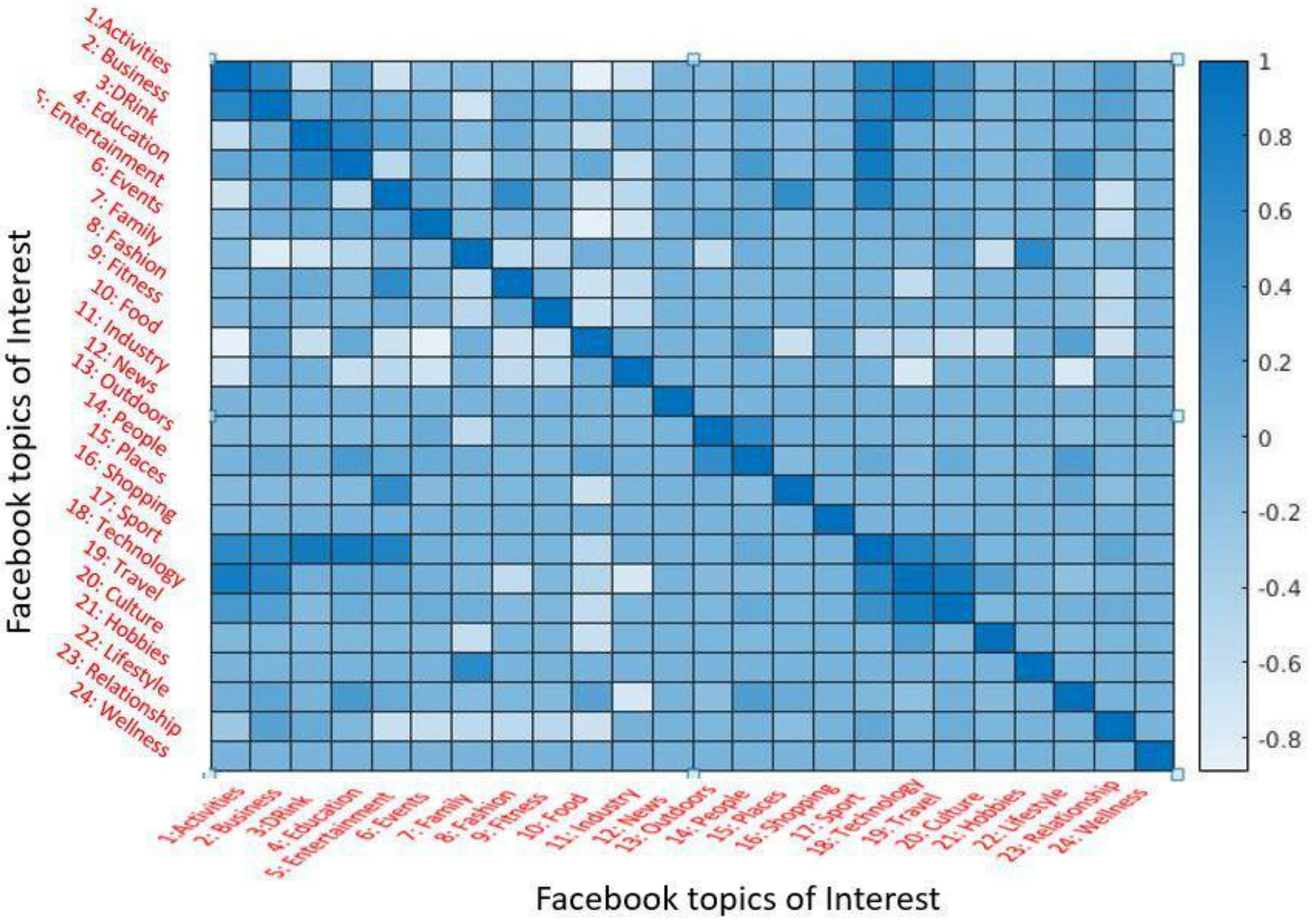

Fig. 12: Pearson Moment Coefficients between various Facebook topics of Interests in table I.

1) Image-level interests distribution study: The main objec-tive of the image-level distribution is to demonstrate the topics partitioning for all users shared images. For that, we apply our annotated test database Sm@rtCityZenDB of 24 classes. Every class contains 10 distinct Facebook users and each user possess 100 diver shared images.

To illustrate this level , we propose a demonstration about some classes chosen randomly that are Sport, Family and Education respectively.

According to the figures 16 ,in which each user possess 50 shared images, we remark that the images topics distribution is articulated around three topics which are Outdoors, Drink, People and Relationship. This distribution, validate the posi-tive correlation between these topics and the fact that a user who is interested by the topic Sport, he/she may share images belong these discovered topics. In fact, the shared images for a user interested by the topic Sport contains concepts belong the super class Outdoors or Drink or People or Relationship in our ontology. Additionally, the interrelationship between Sport and other predicted topics is explained by the fact that the sport practitioner needs the water and that outdoor activity is a good way to put the fun into sport and especially with friends. For the same class, the figure 17 illustrate this distribution level using 100 shared images per user. In fact, with 100 images, we remark that the images distribution has become more detailed with the appearance of new topics with low scores such as Shopping, Places, Entertainement and Fitness and the disappearance of the self-assessed topic for some users. This appearance and disappearance is explained by the diversity of images that can generate vectors with low scores for several topics.

Furthermore, the figure 18 is concerned for the class Fa-mily. In this case, the users' images have as distribution the topics Food, Drink, Shopping, Outdoors and Family. This distribution is explained by the fact that a user who is interested by Family is strongly interested by Food, Drink and Outdoors as the family meets and take pictures, most of the time, on the dining tables and in the gardens. Using 100 shared images, from the figure 19, a new topics have appeared like Sport, Entertainement and a low scores are assigned to the self-assessed topic, compared to the case of 50 images.

In addition, we explain the image level distribution for the topic Education illustrated in figure 19. After observation, we concluded that the users' images express the topics Drink, Sport, culture and Fashion. This distribution validate the high positive correlation between Education and Culture but it is not significant for the Sport and Fashion topics prediction. Applying 100 images per user, the figure 20 illustrate the novel distribution for each user' images.

To conclude, we can assume that in the social network Facebook, each user may share a set of images whose can be related or not related to his/her self-assessed topic of interest.

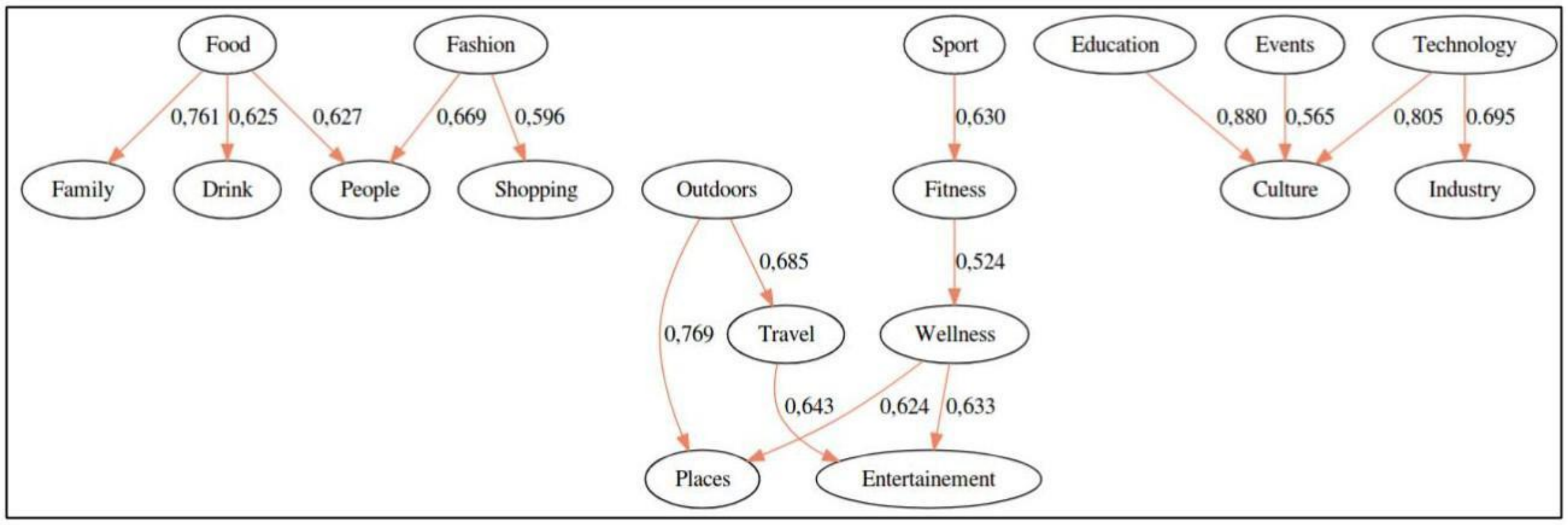

Fig. 13: High correlation between Facebook Topics of Interests

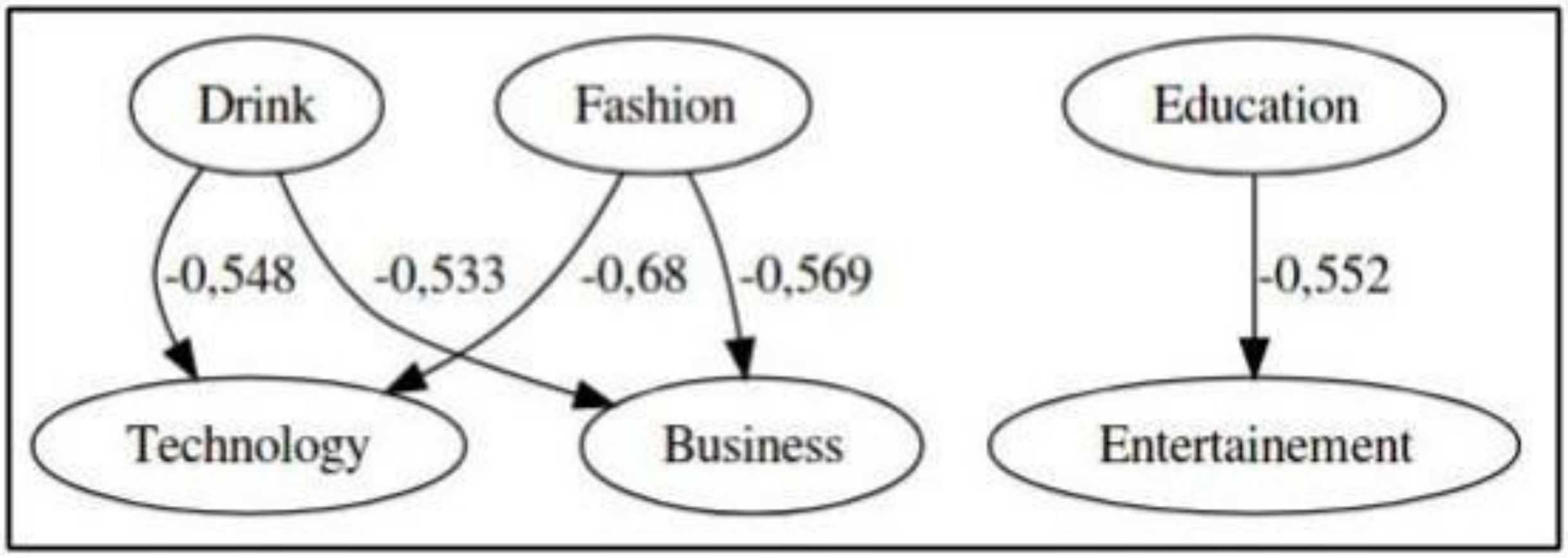

Fig. 14: High negative correlation between Facebook Topics of Interests

With 50 images, the distribution is generally very close to reality with the appearance of the self-assessed with the most important score and with 100 images, this distribution has become more detailed with new topics assigned to that are correlated with self-assessed topic for each user.

2) User-level interests distribution study: As we have already explained in Algorithms 2 and 3, the output matrices of image-level distribution are the input for the user-level. In fact,the interests distribution for a given user, is quite simply the total distribution of all his/her shared images.

In this level, we try to define the confidentiality area of number of users shared images. For that, we used k shared images by each user within the 24 classes. According to each class, we have obtained a confidential area which resulted the target user interests matrix with high score for the self-assessed topic.

The table VI describe the variation of the accuracy measure for each class according to the number of shared images per user. We used the Cumulative match characteristics (CMC) curve to illustrate the evolution of interests prediction rate with the number of shared images and to compare this rate for each class.

Firstly, the figure 22 present the CMC curve for the classes Actitivies, Business, Drink, Education, Entertainement and Events to represent their accuracies evolution. In fact, we remark that Activities and Business have the same accuracy value for 5 or 10 or 50 shared images and the value 0 for 100 images. Some classes possess a high accuracy value for a specific number of images that are Drink and Entertainement which have accuracy 1 for 10 images and 5/10/50 images respectively. For Education and Events classes, it present the stability of accuracy value for 5 or 10 or 50 images and 5 or 10 images respectively. Secondly in the figure 23, we show the evolution of accuracies values with the number of shared images for Family, Fashion, Fitness, Food, Industry and News classes. We observe, that Fashion and News classes have the same accuracy value for 5 or 10 or 50 images which are 1 and 0 respectively and the same accuracies for 75 and 100 images. Family class obtain an accuracy of 1 with 50 shared images and 0 for 100 images, Food class posses an accuracy of 1 with 5 images and Industry class obtain an accuracy of 1 for 5 and 10 shared images per user. The class Fitness achieve an accuracy of 1 for 50 shared images. Then, the figure 24 illustrate the CMC curve for Outdoors, People, Places, Shopping, Sport and Technology classes. Through this figure, we observe that Outdoors and Shopping classes get a stability of accuracy value for 5 or 10 or 50 shared images which are 1 and 0.5 respectively and 0 for 100 shared images The classes Places, Sport and Technology achieve an accuracy of 1 with 10 shared images per user. However, the class Peoples obtain the high accuracy of 0.7 for 50 shared images. Finally, the figure 25 present the Travel, Culture, Hobies, Lifestyle, Relationship and Wellness classes accuracies with the change of number of shared images per user. The class Travel possess a high accuracy of 1 which only 5 shared image, although, the classes Culture, Relationship and Wellness obtain the high accuracy for 50 images. The class Lifestyle achieve an accuracy of 0 for the five case studies. The class Hobies get the same accuracy of 1 for 10 and 50 shared images and 0 for 75 or 100 shared images.

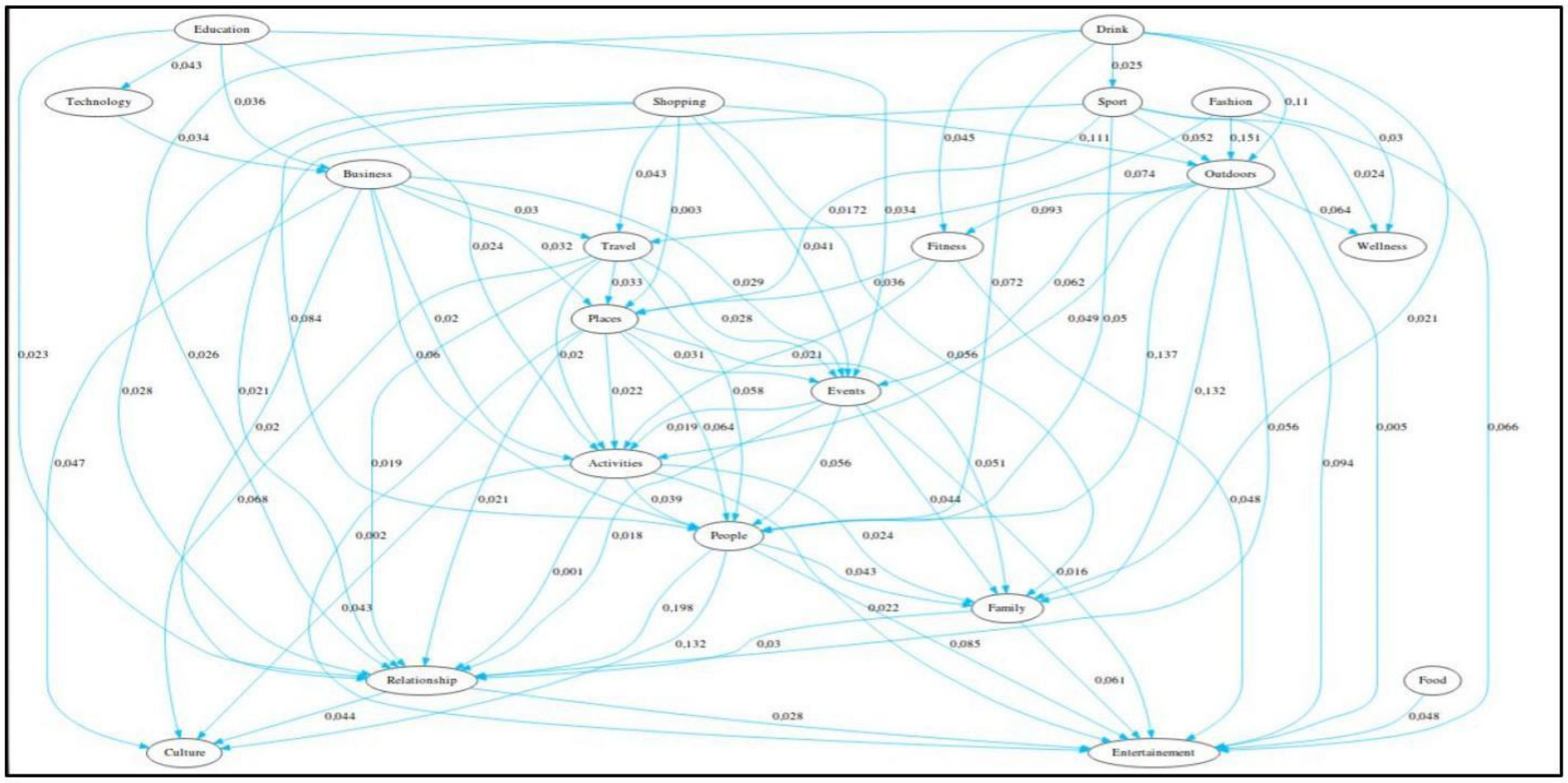

Fig. 15: Medium correlation between Facebook Topics of Interests

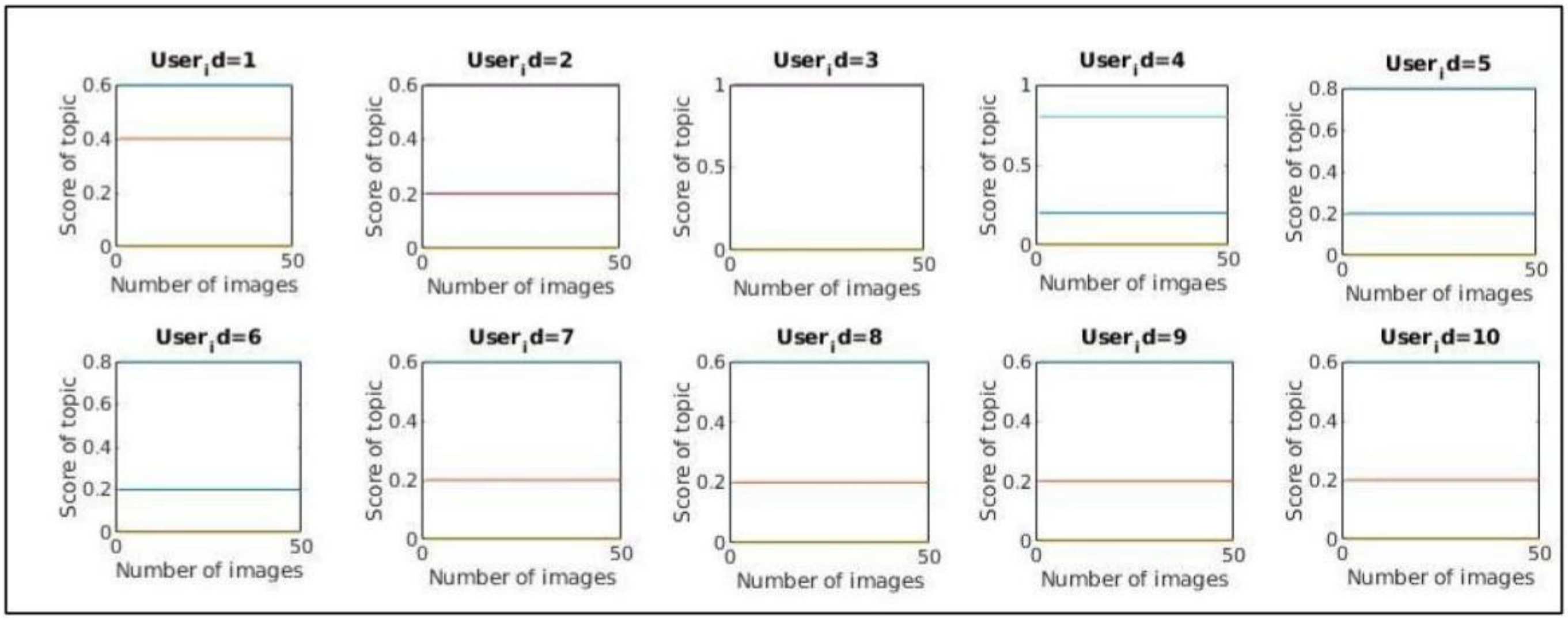

Fig. 16: Image-Level distribution for Sport class with 50 images per user: (user1)Outdoors:0.6,Drink:0.4 (user2)People:0.6, Drink:0.4 (user3)People:1 (user4)Outdoors:0.8, Relationship:0.2 (user5)Outdoors:0.8, Relationship:0.2 (user6)Outdoors:0.8, Relations-hip:0.2 (user7)Outdoors:0.6, Drink:0.2, People:0.2 (user8)Outdoors:0.6, Drink:0.2, People:0.2 (user9)Outdoors:0.6, Drink:0.2, People:0.2 (user10)Outdoors:0.6, Drink:0.2, People:0.2 .

To visualize the performance of our prediction algorithm , we illustrate, in figures 26, 27, 28, 29 and 30, the co-ocurrence matrices specific for our five case studies. In each matrix, the row represents the instance in a predicted class while each column represents the instance in a original class.

The table VII demonstrate the variation of our system performance according to the number of shared images by users. We can explain this variation by the fact that each user' topics of interests distribution consists of three layers: Start term with 5 images, Middle term with 10 images , Long term with 50 images, Very long term with 75 images and Extreme

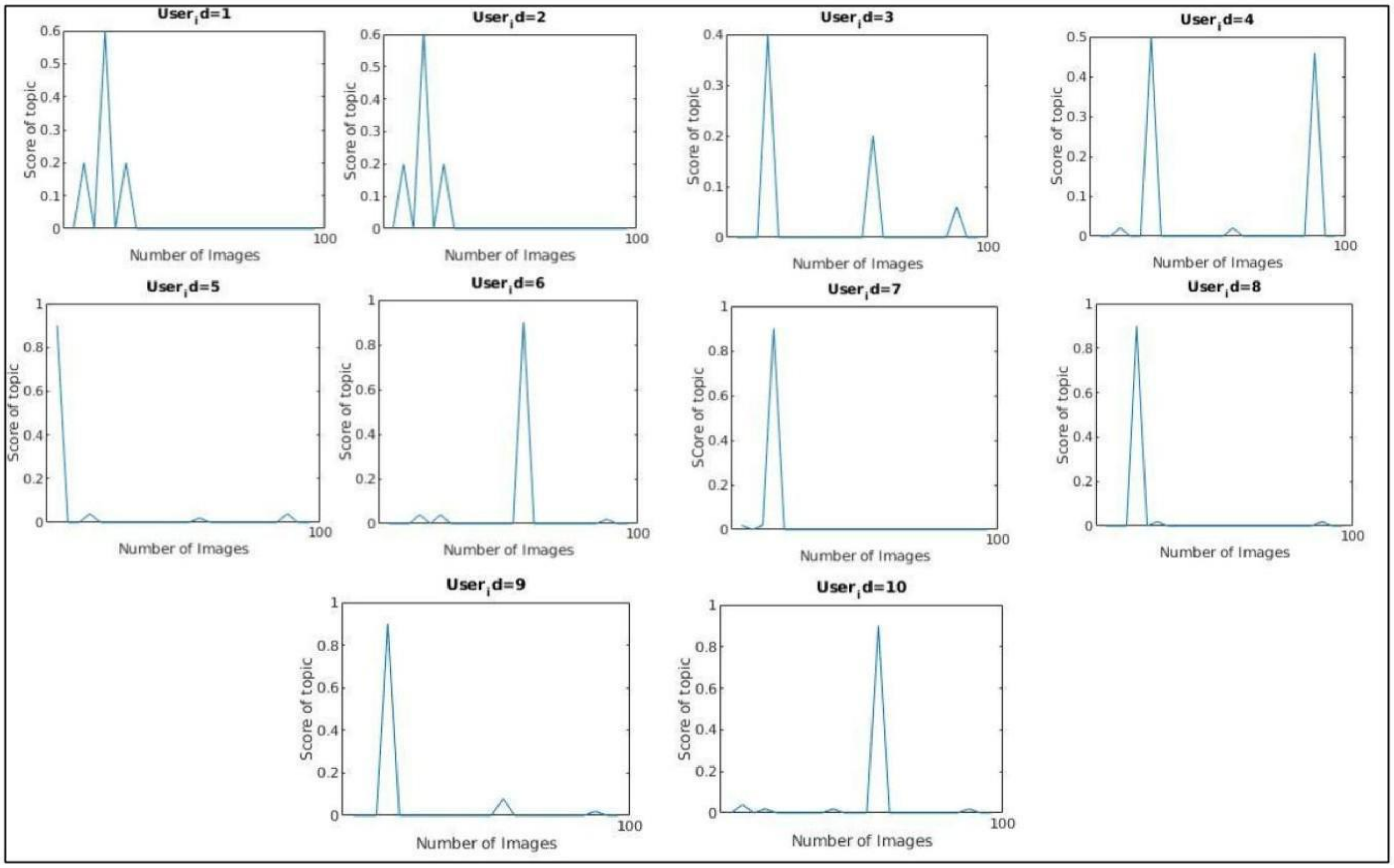

Fig. 17: Image-Level distribution for Sport class with 100 images per user: (user1)Drink:0.2, Shopping:0.6, Outdoors:0.2 (user2)Drink:0.3, Shopping:0.5, Outdoors:0.2 (user3)Shopping:0.4, Places:0.2, Entertainement:0.6 (user4)Fashion:0.02, Sport:0.5, Places:0.02, Entertainement:0.46 (user5)Drink:0.9, Fashion:0.04, Places:0.02, Entertainement:0.04 (user6)Shopping:0.04, Outdoors:0.04, Places:0.9, Entertainement:0.02 (user7)Food:0.02, Fashion:0.02, Shopping:0.9 (user8)Shopping:0.9, Outdoors:0.02, Entertainement:0.02(user9)Shopping:0.9, Places:0.08, Entertainement:0.02 (user10)Drink:0.04, Shopping:0.02, Fitness:0.02, Places:0.9.

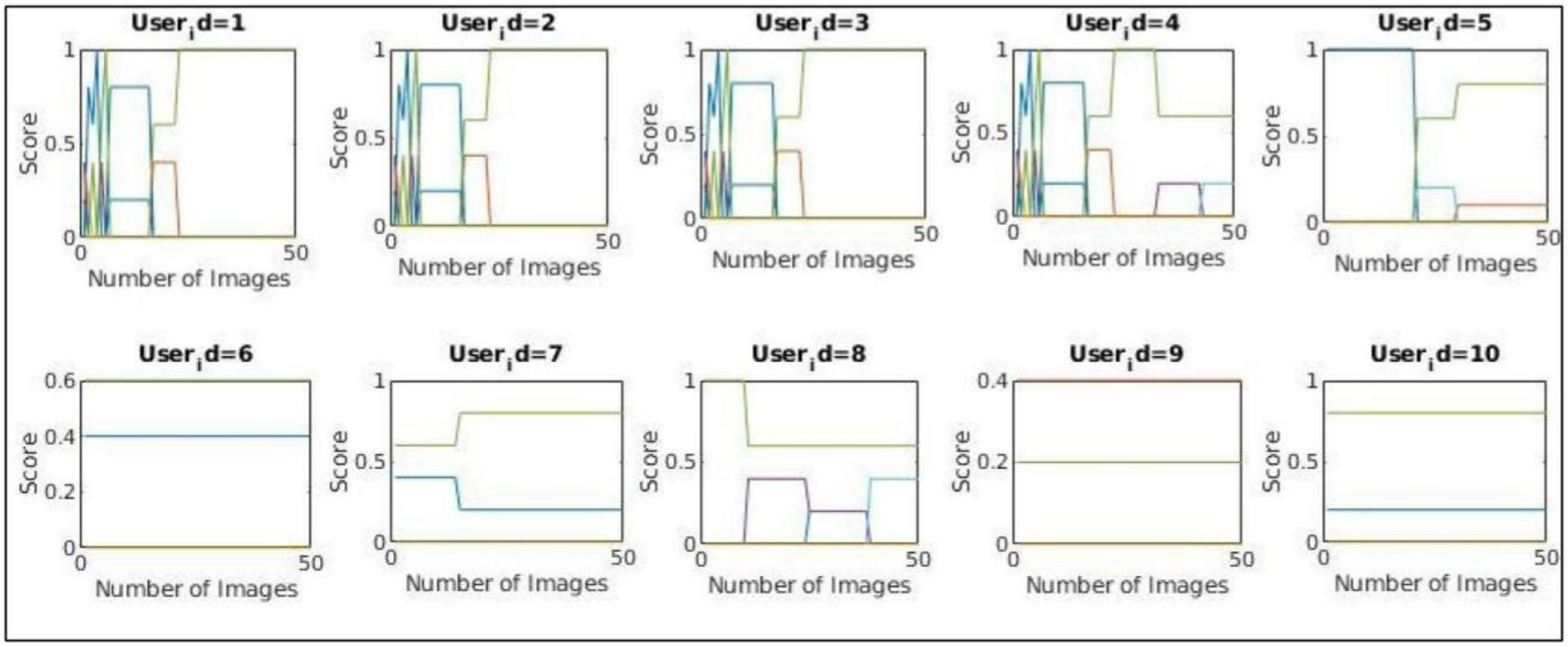

Fig. 18: Image-Level distribution for Family class with 50 images per user: (user1)Drink:0.12, Fashion:0.02, Shopping:0.02, Sport:0.02, Family:0.76 (user2)Fashion:0.02, Sport:0.02, Family:0.96 (user3)Food:0.03, Drink:0.02, Shopping:0.02, Family:0.86, Culture:0.02 (user4)Food:0.1, Family:0.9 (user5)Shopping:0.02,Family:0.98 (user6)Food:0.02, Sport:0.02, Family:0.96 (user7)Food:0.02, Fashion:0.02, Family: 0.96 (user8)Food:0.02, Family:0.98 (user9)Food:0.04, Sport:0.02, Family: 0.94 (user10)Food:0.06, Shopping:0.02, Family: 0.94

term with 100 images.

The start term present the sharing of the first 5 images

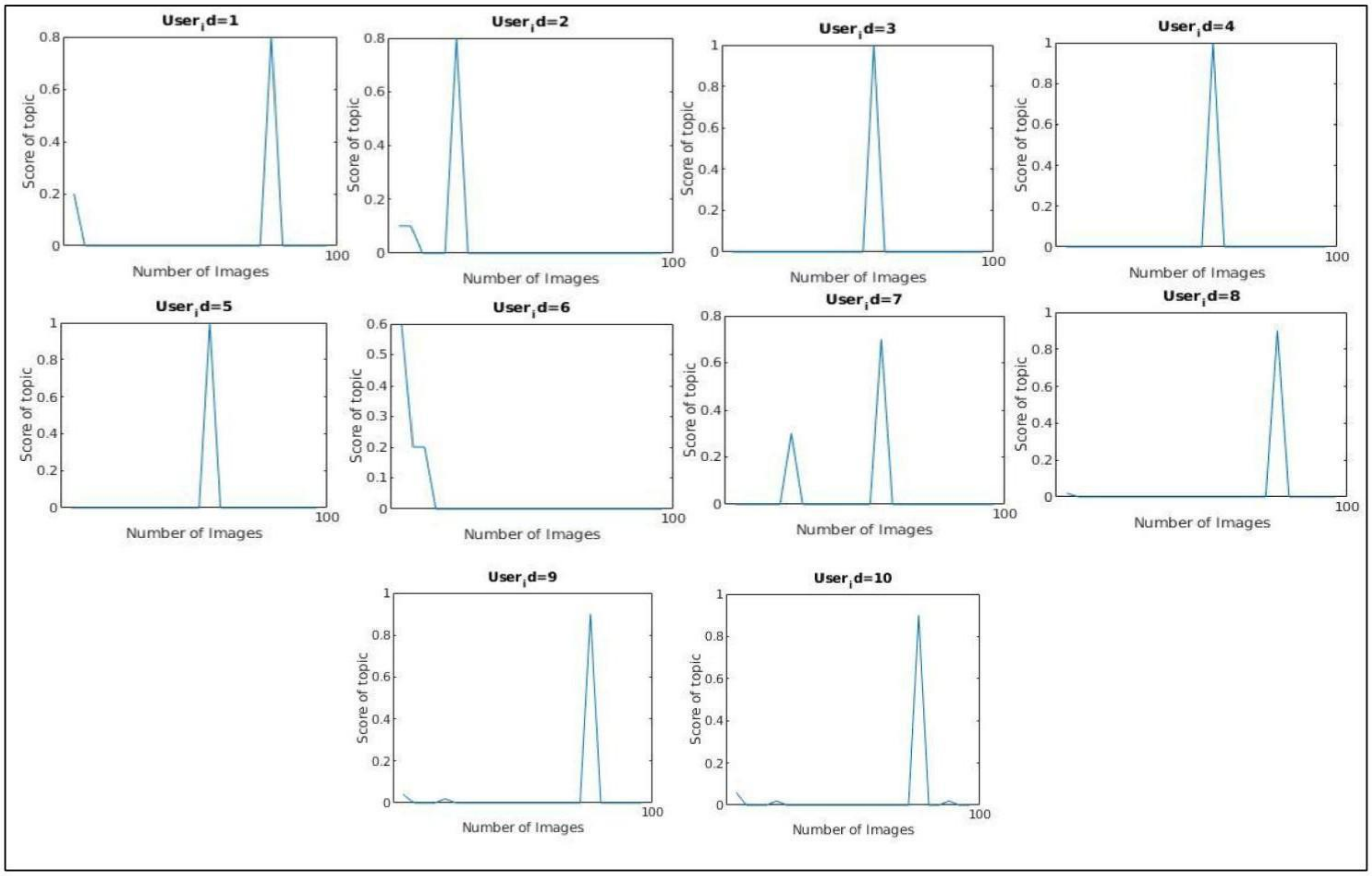

Fig. 19: Image-Level distribution for Family class with 100 images per user: (user1)Food:0.2, Family:0.8(user2)Food:0.1, Drink:0.1, Outdoors:0.8 (user3)Places:1(user4)Places:1(user5)Places:1(user6)Food:0.6, Drink:0.2, Fashion:0.2 (user7)Outdoors:0.3, Places:0.7 (user8)Food:0.01, Family:0.9, Drink:0.09 (user9)Food:0.04,Sport:0.02, Family:0.9, Entertainement:0.02(user10)Food:0.06, Sport:0.02, Fa-mily:0.9, Entertainement:0.02.

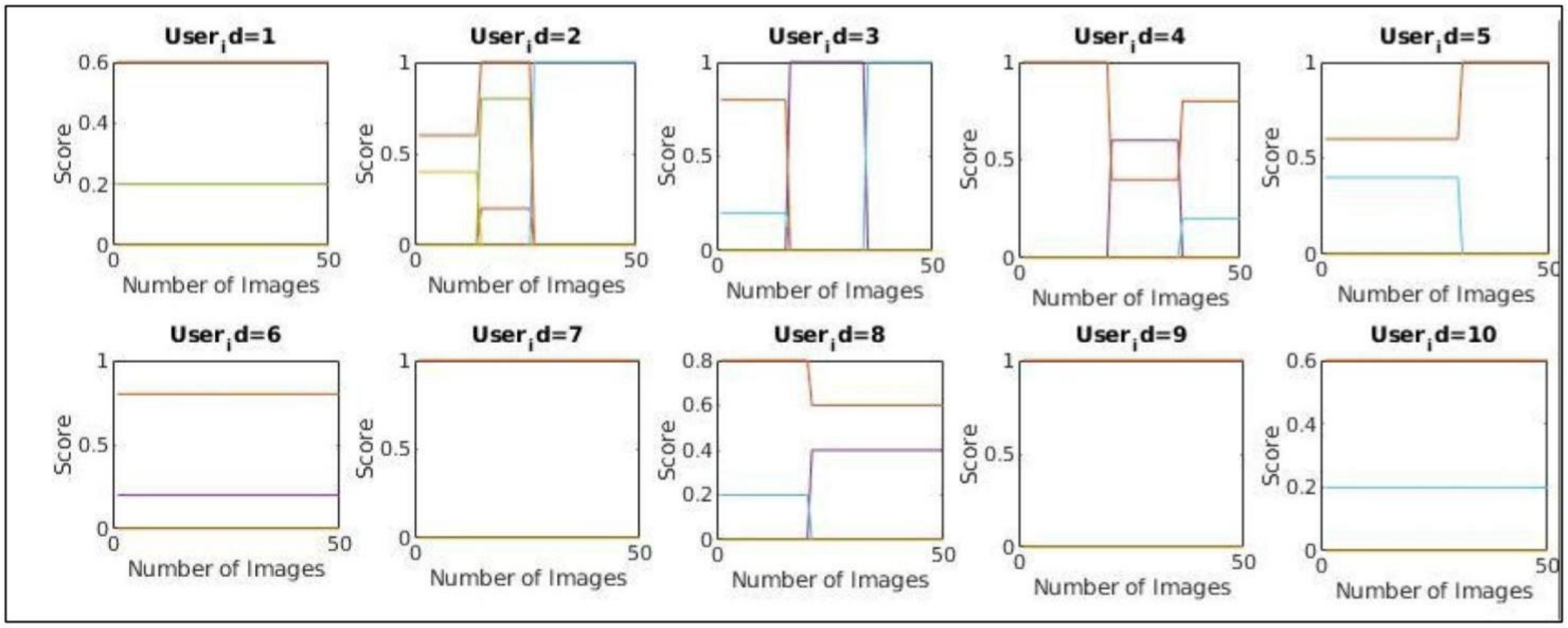

Fig. 20: Image-Level distribution for Education class with 50 images per user: (user1)Drink:0.02, Fashion:0.04, Education:0.92, Culture: 0.02 (user2)Education:1 (user3) Education:0.6, Drink:0.4 (user4)Education:0.6, Drink:0.4 (user5)Education:0.6, Drink:0.4 (user6)Education:0.6, Culture:0.4 (user7)Education:0.8, Drink:0.2 (user8)Education:0.6, Culture:0.4 (user9)Education:0.6, Culture:0.4 (user10)Education:0.8, Cul-ture:0.2

whose each user chooses images indicating their self-assessed topic of interest.

For the middle term, the same user may be influenced by other topics and can share some images that can interrupt

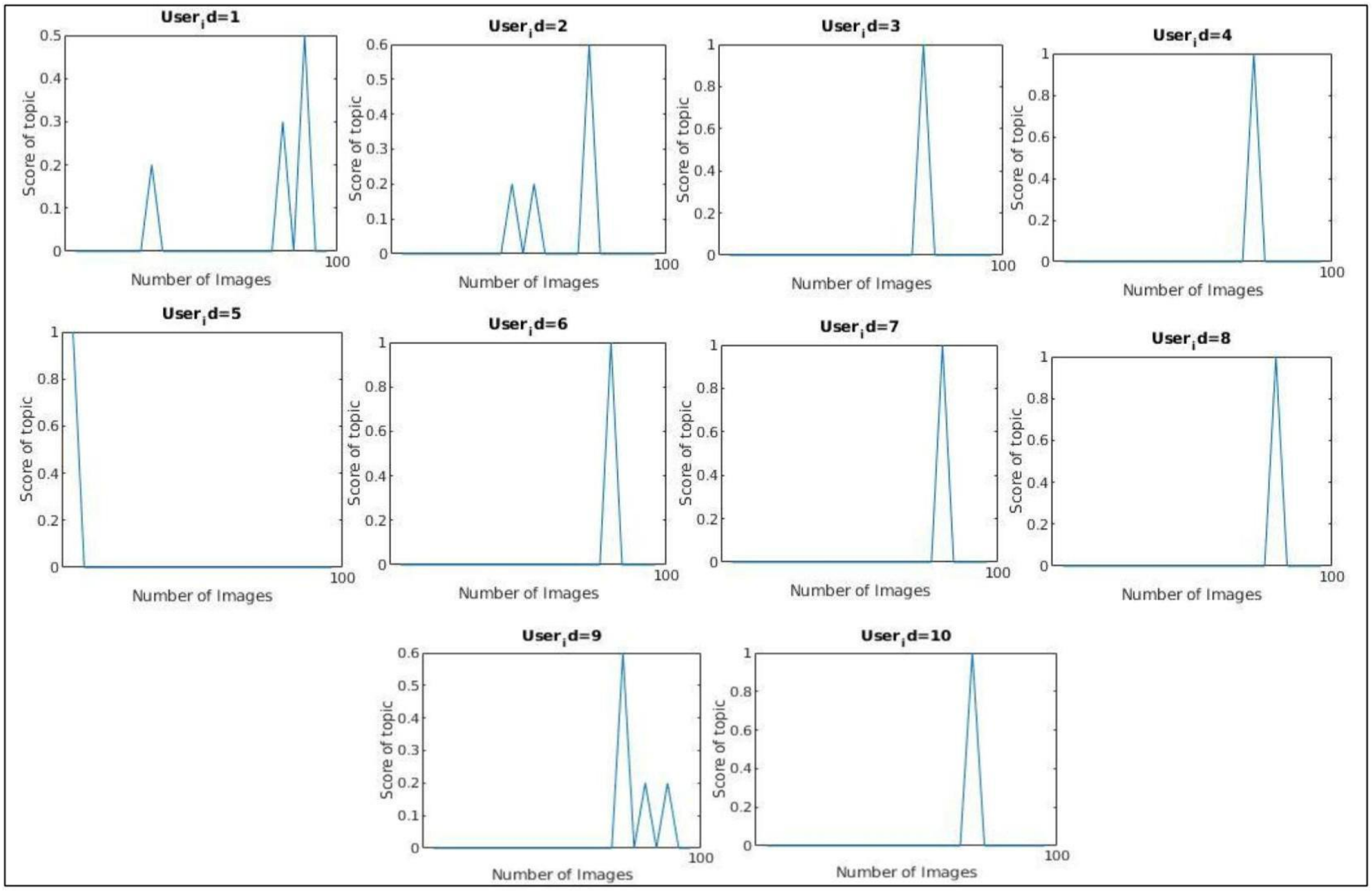

Fig. 21: Image-Level distribution for Education class with 100 images per user: (user1)Fitness:0.2, Relationship:0.3, Culture:0.5 (user2)Business:0.2, Travel:0.2, People:0.6(user3) People:1(user4)People:1(user5)Culture:1 (user6)Culture:1(user7)People:1(user8)People:1 (user9)People:0.6, Culture:0.2, Entertainement:0.2(user10)People1.

TABLE VI: The accuracy measure variation for each class with k shared images: k=5 / k=10 / k=50 / k=75 / k=100

| Classe / Number of Images | 5 | 10 | 50 | 75 | 100 |
|---|---|---|---|---|---|
| Activities | 0.3 | 0.3 | 0.3 | 0 | 0 |
| Business | 0.7 | 0.7 | 0.7 | 0.4 | 0 |
| Drink | 0.6 | 1 | 0.9 | 0.4 | 0.4 |
| Education | 1 | 1 | 1 | 0.4 | 0 |
| Entertainement | 0.9 | 0.6 | 1 | 0.7 | 0 |
| Events | 1 | 1 | 0.8 | 0.3 | 0.3 |
| Family | 0.4 | 0.8 | 1 | 0.6 | 0 |
| Fashion | 1 | 1 | 1 | 0.7 | 0 |
| Fitness | 0.4 | 0.6 | 1 | 0.5 | 0 |
| Food | 1 | 0.8 | 0.9 | 0.6 | 0 |
| Industry | 1 | 1 | 0 | 0.2 | 0 |
| News | 0 | 0 | 0 | 0.5 | 0 |
| Outdoors | 1 | 1 | 1 | 0.5 | 0 |
| People | 0.4 | 0.6 | 0.7 | 0.5 | 0 |
| Places | 0.7 | 1 | 0.4 | 0.5 | 0 |
| Shopping | 0.5 | 0.5 | 0.5 | 0.3 | 0 |
| Sport | 0.5 | 1 | 0.5 | 0.3 | 0 |
| Technology | 0.6 | 1 | 0.7 | 0.3 | 0 |
| Travel | 1 | 0.2 | 0.4 | 0.2 | 0 |
| Culture | 0.6 | 0.4 | 1 | 0.8 | 0 |
| Hobies | 0.5 | 1 | 1 | 0 | 0 |
| Lifestyle | 0 | 0 | 0 | 0 | 0 |
| Relationship | 0.1 | 0 | 0.7 | 0.5 | 0 |
| Wellness | 0 | 0 | 0.5 | 0.2 | 0 |

our classification which explains the decrease of system performance from 0.85 to 0.75.

In a long term, after being biased in the middle term, the user settle back into her/his self-assessed topic of interest and our system predict the correct target class with 50 shared images to obtain an accuracy of 0.95.

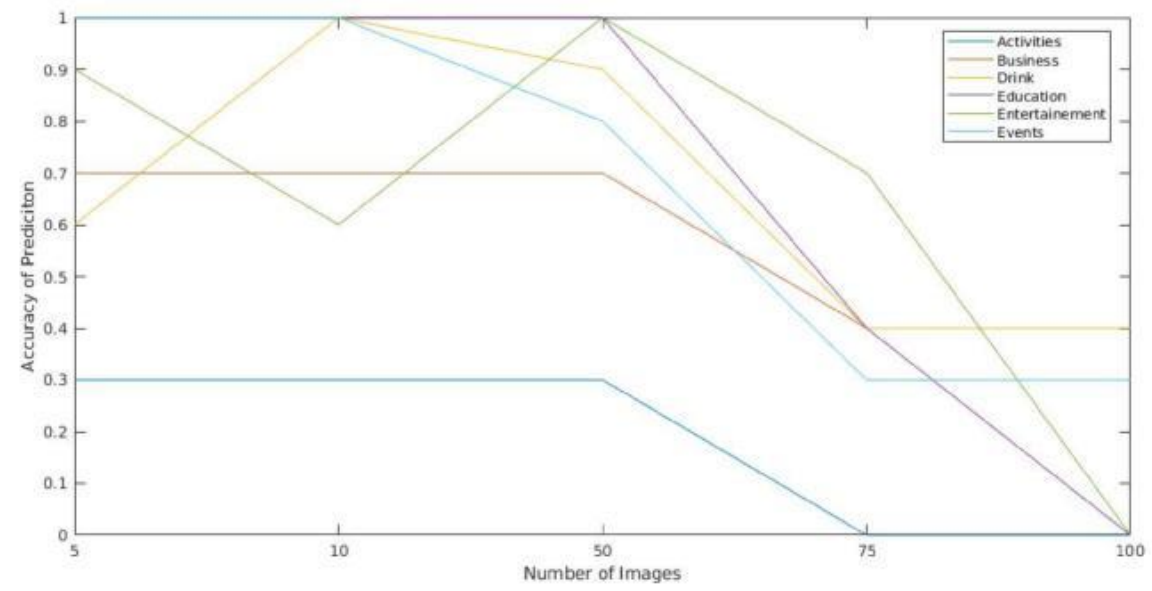

Fig. 22: Cumulative match characteristics (CMC) curve for Activities, Business, Drink, Education, Entertainment and Events classes

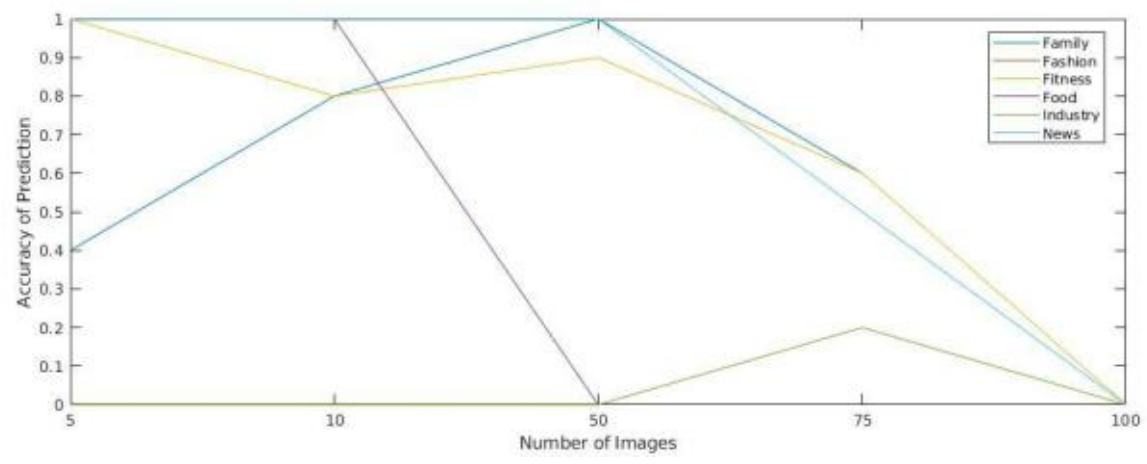

Fig. 23: Cumulative match characteristics (CMC) curve for Family, Fashion, Fitness, Food, Industry and News classes

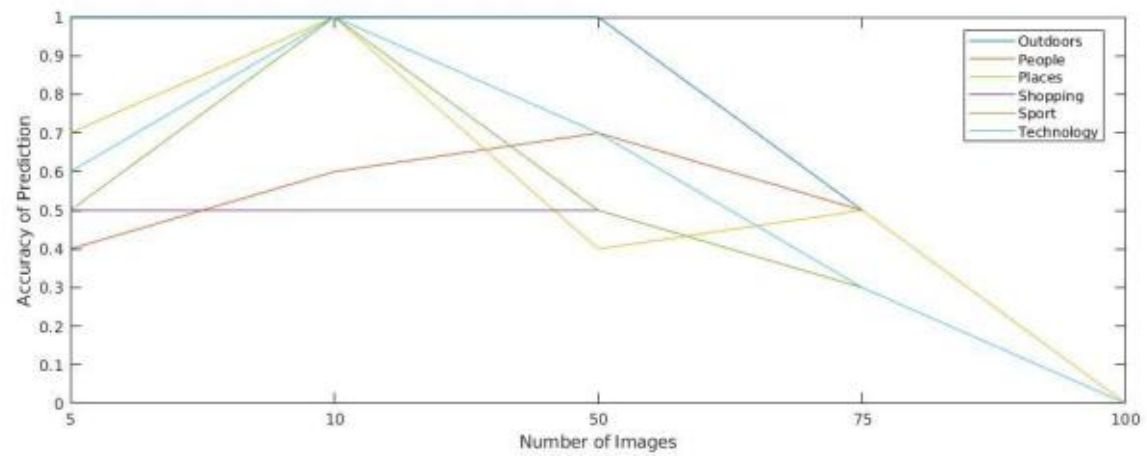

Fig. 24: Cumulative match characteristics (CMC) curve for Outdoors, People, Places, Shopping, Sport, Technology classes

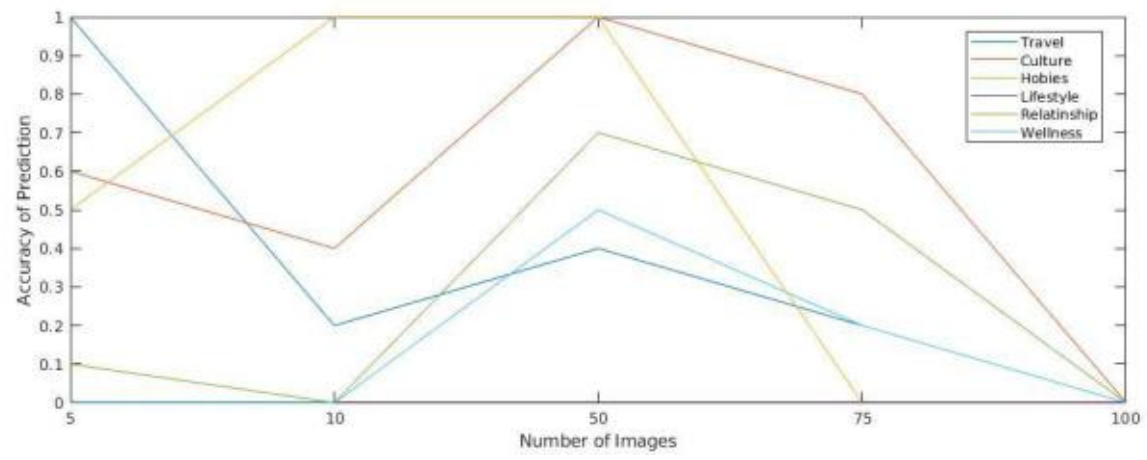

Fig. 25: Cumulative match characteristics (CMC) curve for Outdoors, People, Places, Shopping, Sport, Technology classes

TABLE VII: The variation of DeepVisInterests performance accor-ding to number of shared images per user

| Number of shared Images | Accuracy |
|---|---|
| 5 images | 0.85 |
| 10 images | 0.75 |
| 50 images | 0.95 |
| 75 images | 0.80 |
| 100 images | 0.65 |

In a very long term, the user keep a stability with a slight disturbance of its distribution obtained in the long term. For that, our system present shows a slight decrease in performance with an accuracy of 0.80.

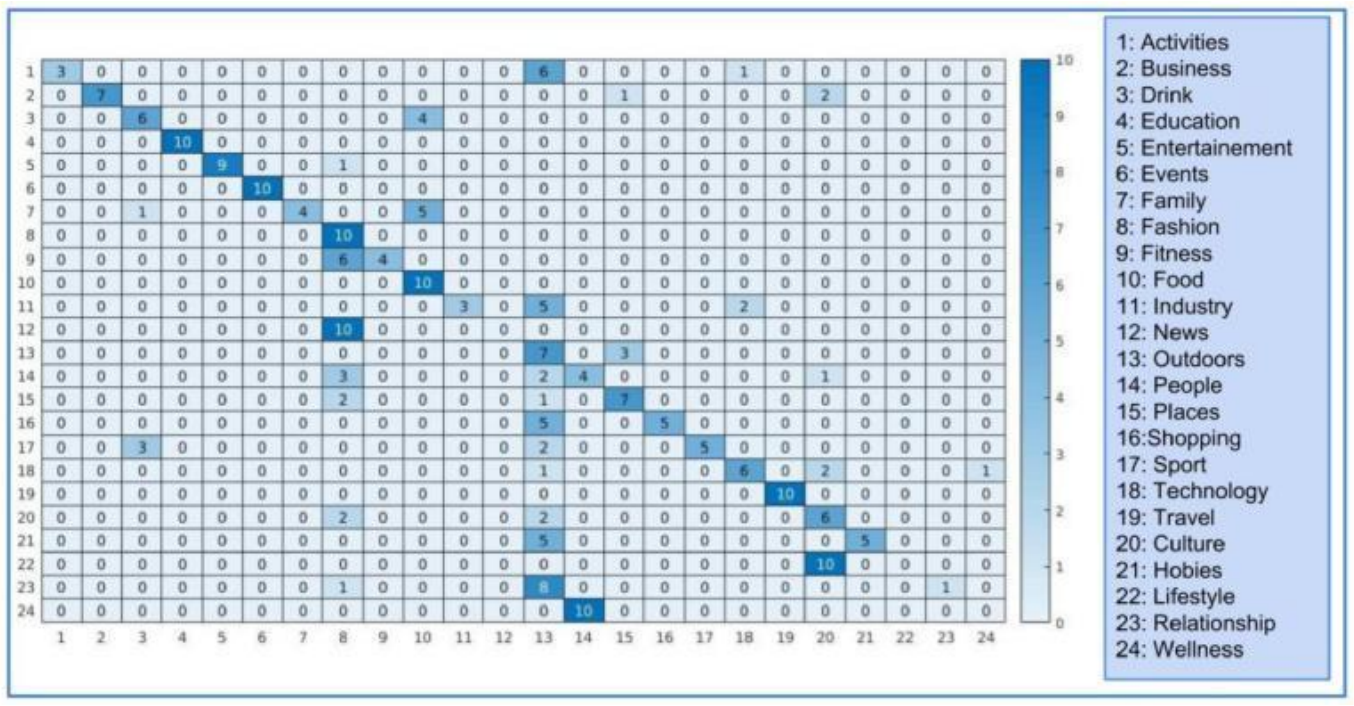

Fig. 26: Co-ocurrence matrix of users interest prediction with 5 shared images per user

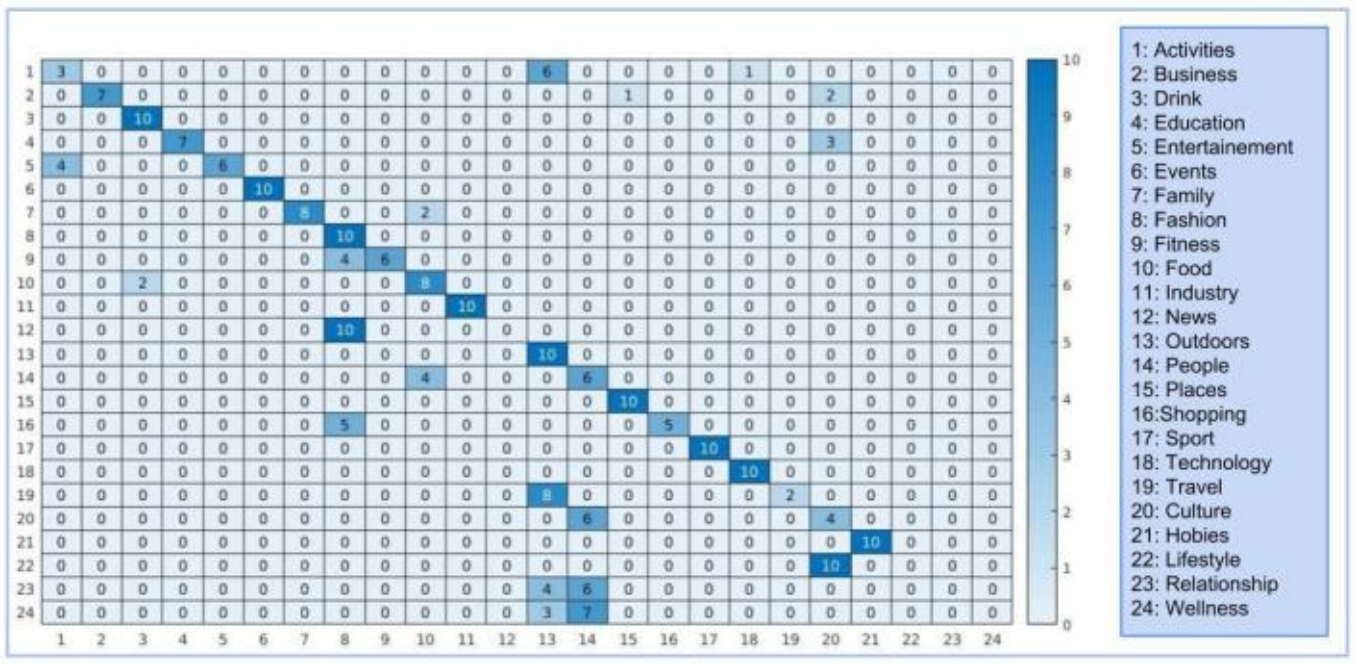

Fig. 27: Co-ocurrence matrix of users interest prediction with 10 shared images per user

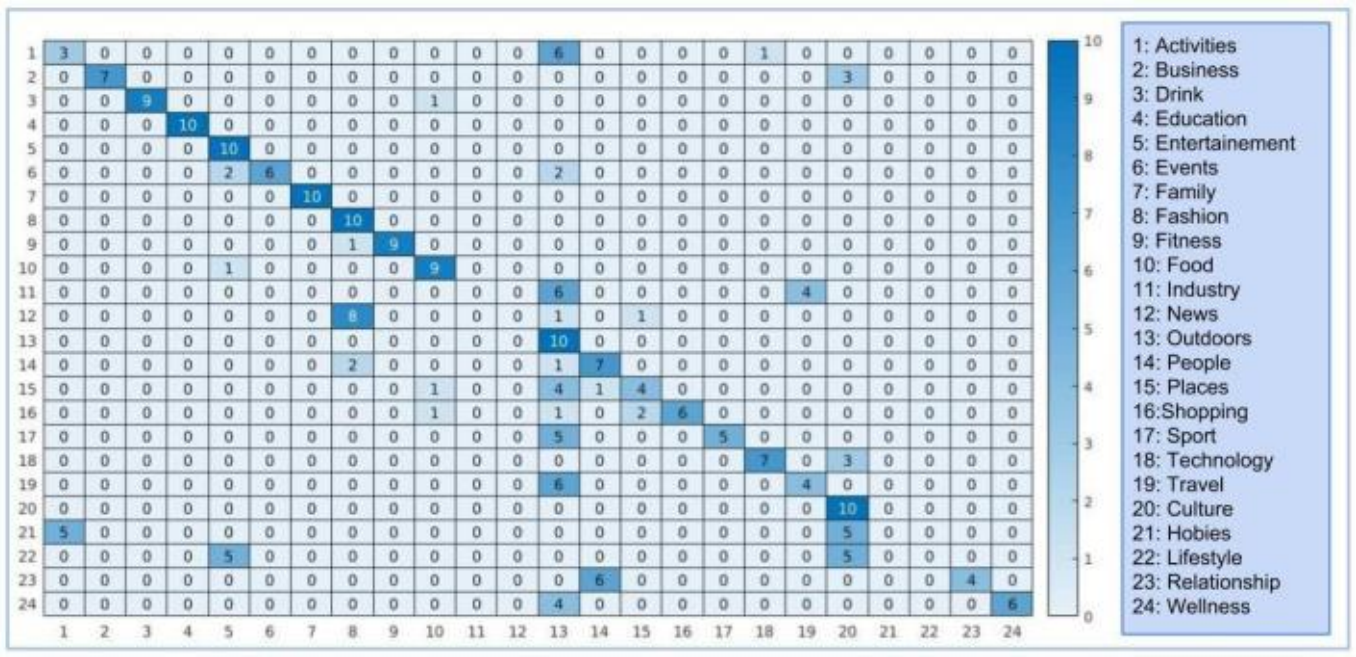

Fig. 28: Co-ocurrence matrix of users interest prediction with 50 shared images per user

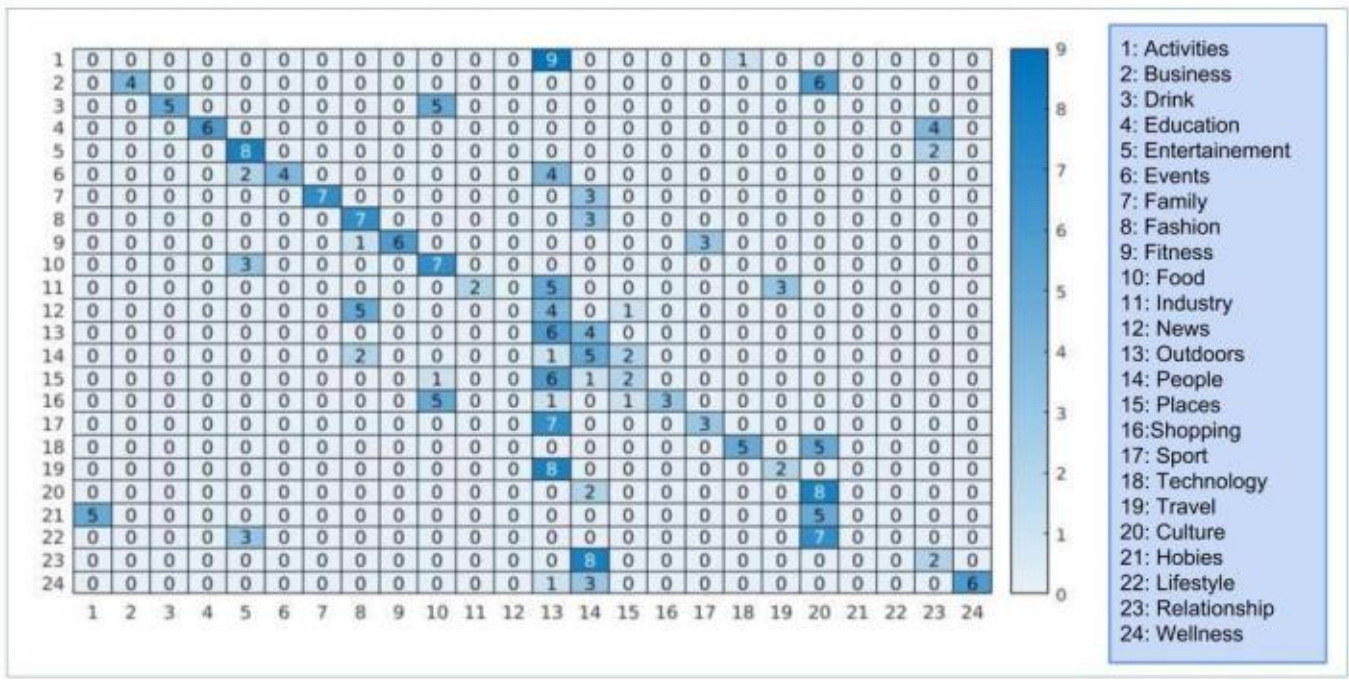

Fig. 29: Co-ocurrence matrix of users interest prediction with 75 shared images per user

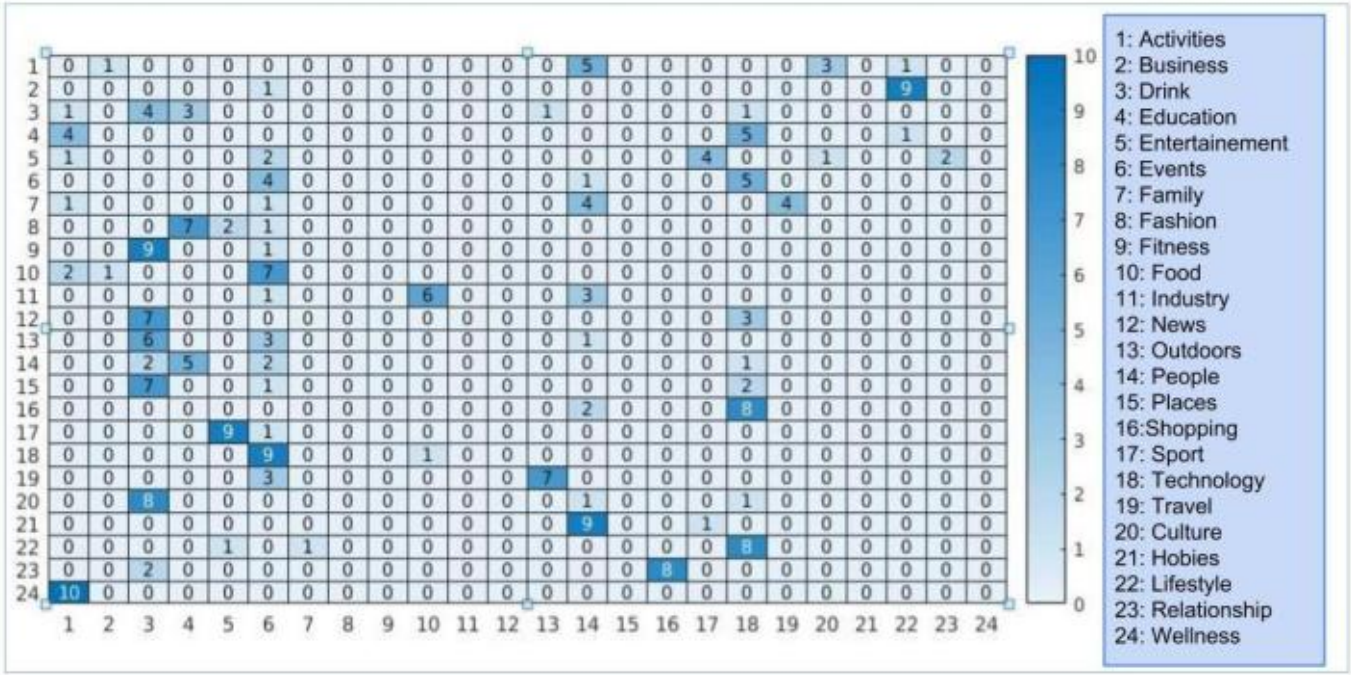

Fig. 30: Co-ocurrence matrix of users interest prediction with 100 shared images per user

In the extreme long term, our system performance under-goes a very remarkable decrease with an accuracy of 0.65 which validates that beyond 50 images the distribution of the centres of interest undergoes a disturbance by the diversity of the images which influences negatively on the self-assessed topic.

To evaluate the performance of our algorithm, we apply two divers criteria. 1) Precision, is a popular measure to represent the fraction of relevant classes among the predicted classes.

$$Precision = \frac{TP}{TP + FP} \quad (8)$$

2) Recall is the fraction of relevant classes that have been retrieved over the total amount of relevant classes.

$$Recall = \frac{TP}{TP + FN} \quad (9)$$

The figures 31 and 32 illustrate the precision and recall for our prediction algorithm for the three case studies, respectively.

## C. CNN architectures for users interests prediction

To evaluate our proposed framework, we hence deliver a continuously growing set of pre-trained models with famous architectures for the Caffe Framework to extract the visual features from social images in the test database. One focus on our work is about the depth of the CNN, which affect the ability of capability of convolution. Thus, we make use of four different CNN architectures that are AlexNet [22], LeNet [24], VGG19 [36] and GoogleNet [39] to highlight a comparative study between these performances illustrated on table VIII in terms of test accuracy performance for prediction.

TABLE VIII: Users Interests prediction Accuracy based on CNN architectures

| Architecture | Accuracy |
|---|---|
| LeNet | 76.53 |
| AlexNet | 81.03 |
| VGG19 | 91.32 |
| GoogleNet | 93.25 |

Despite the fact that our results have almost the same average values within statistical fluctuations, we assume that GoogleNet architecture accommodates additional advantages. In fact, we focus on this architecture as they provide marginally more appropriate results view that going deeper allow the network to disentangle progressively more abstract concepts.

## D. Comparison with State-of-the art

The advances in the number of social networks users, in turn, has the amount of social content. Different studies are being conducted based on this area, with topics of interests prediction and classification being one of the topics based on social networks.
However, each work try to classify the social interests based on a set of predefined categories or topics using a specific type of social data. To show the originality and the efficiency of our proposed framework DeepVisInterests, we compare our results with others works, which their goals are to perform the topics of interests prediction and distribution.
From table IX, we can assume that our method outperforms the literature methods in several levels.

For the social data sources level, the social networks are a major database application area, but currently there are few benchmarks. [2]. In this context, we can valorize our framework by choosing Facebook as source of data among other social networks. The social data collection from Facebook requires the implementation of specific application. However, we can only collect data from users accounts who are registered in our developed Facebook application [23]. But other social networks like Pinterest, used practically by the state-of-the art methods [34], [49] and [47], apply the crawling method to create a social database based on a public and available APIs.
Beside the data level and for the visual features extraction, we used four CNN architectures for object recognition to highlight this module compared to other works.

# V. Discussion

In this part of discussion, we will try to discuss the performance of our proposed framework based on the obtained results as well as the used architectures to obtain it and we will try to give some explanation that is intuitive to know as to why an image is misclassified.
We examine our proposed framework by illustrating the users interests distribution on the tweenty-four topics of interests according to the number of shared images by them.

Firstly and in order to find out if our system is globally significant and to compare our several condidate models to discuss the ideal number of shared images used to obtain the best user-level topics of interests distribution, the figure 33 illustrate the Receiver Operating Characteristic (ROC) for the tweety-four used classes which presents the proportion of true positives based on the proportion of false positives. This curve determine the confidential area for our prediction. We remark that the model with 50 images is always above the other four models with 5 or 10 or 75 or 100 images. In fact, we validate that, although 5 images produces better ROC values for higher thresholds, 50 images is usually better at distinguishing the bad radar return from the good ones. The ROC curve for 100 images is generally lower than the other four ROC curves which indicates worse in-sample performance than the other two classifiers methods. For that we can assume that the model of 50 shared images per user will be the best performing given the best user-level topics of interests distribution.

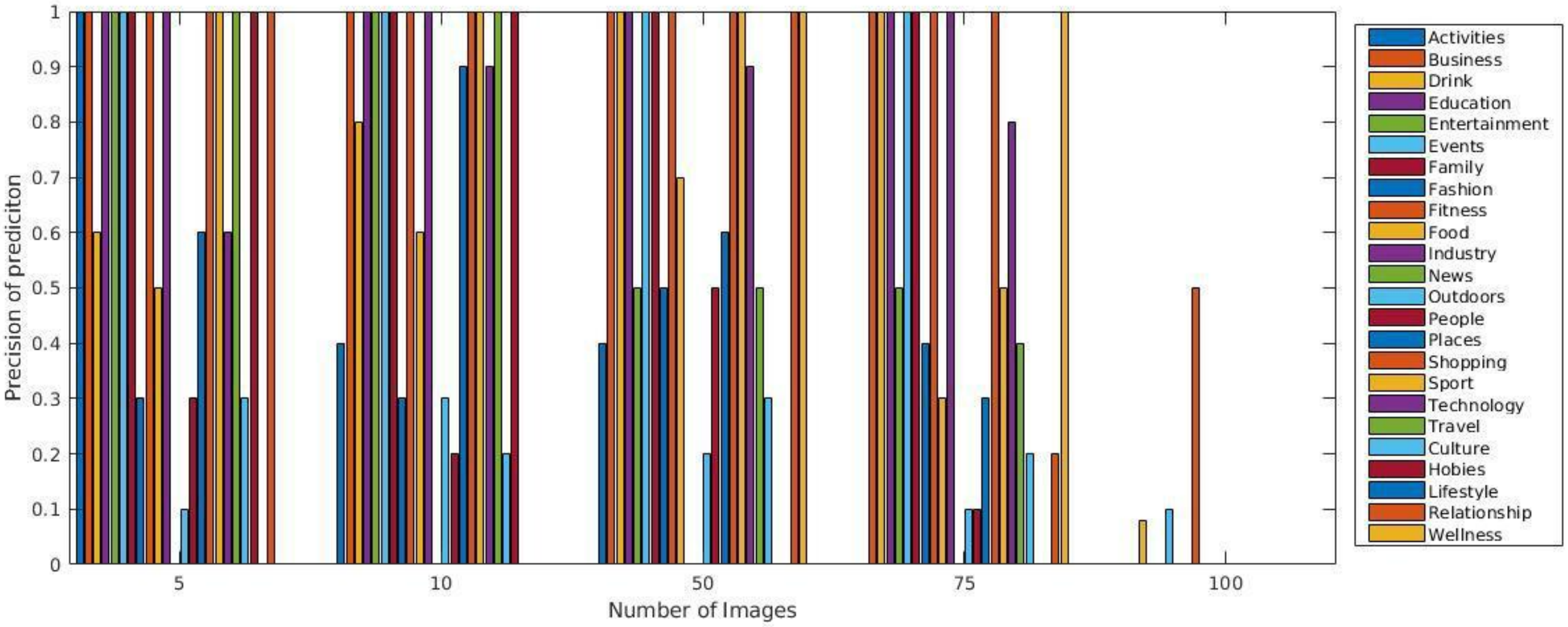

Fig. 31: Precision measure of DeepVisInterests framework

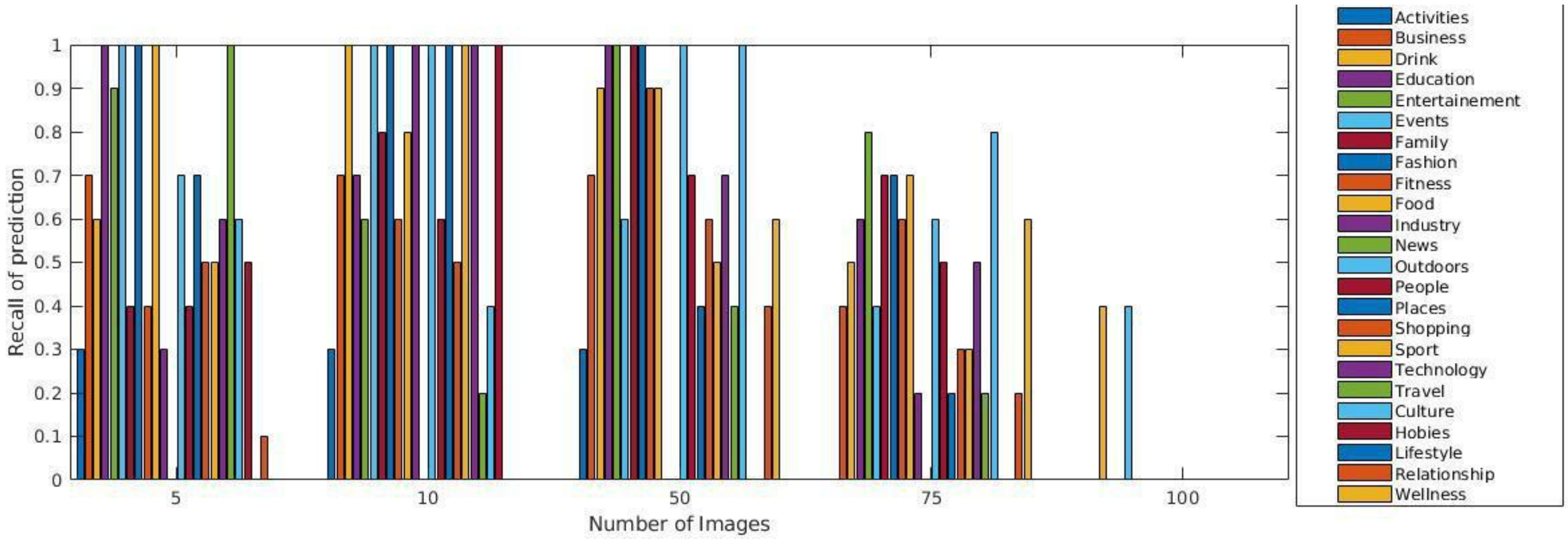

Fig. 32: Recall measure of DeepVisInterests framework

TABLE IX: Comparison between DeepVisInterests and State-of-the art

| Approach | Social Network | Number of images | Features | Number of classes | Accuracy |
|---|---|---|---|---|---|
| [49] | Pinterest | 300 | AlexNet | 7 | 0,43 |
| [34] | Pinterest | 800 | AlexNet and VGG19 | 5 | 0,68 |
| [47] | Pinterest | 20,500 | Siamese | 2 | 0,399 |
| DeepVisInterest | Facebook | 24,000 | Lenet, AlexNet, VGG19 and GoogleNet | 24 | 0,80 |

In addition, we can discuss the obtained results by eva-luating the performance of each architecture used for visual features extraction.

From the table VIII , we fond that GoogleNet architecture is the most for users interests prediction with a high accuracy in the five topics of interests. Yet, each CNN architecture contains two separate modules that are the feature extractor and the classification modules.The performance of any CNN architecture is related to the feature extractor module parame-ters, especially the number and size of filters and layers. First of all, LeNet is one of the first convolutional neural network designed especially to classify handwritten digit. It has 2 convoutional layers, 2 sampling layers, 2 hidden layers and 1 output layer. For the users interests prediction task, LeNet has a low classification accuracy due to the change from hand written digits classification to complex scene images classification.

Secondly, AlexNet is one of the deep convolutional neural network to deal with complex scene classification task. Alex-Net has 5 convoluional layers, 3 sub sampling layers and 3 fully connected layers. It use a set of filters with size of 11 11,

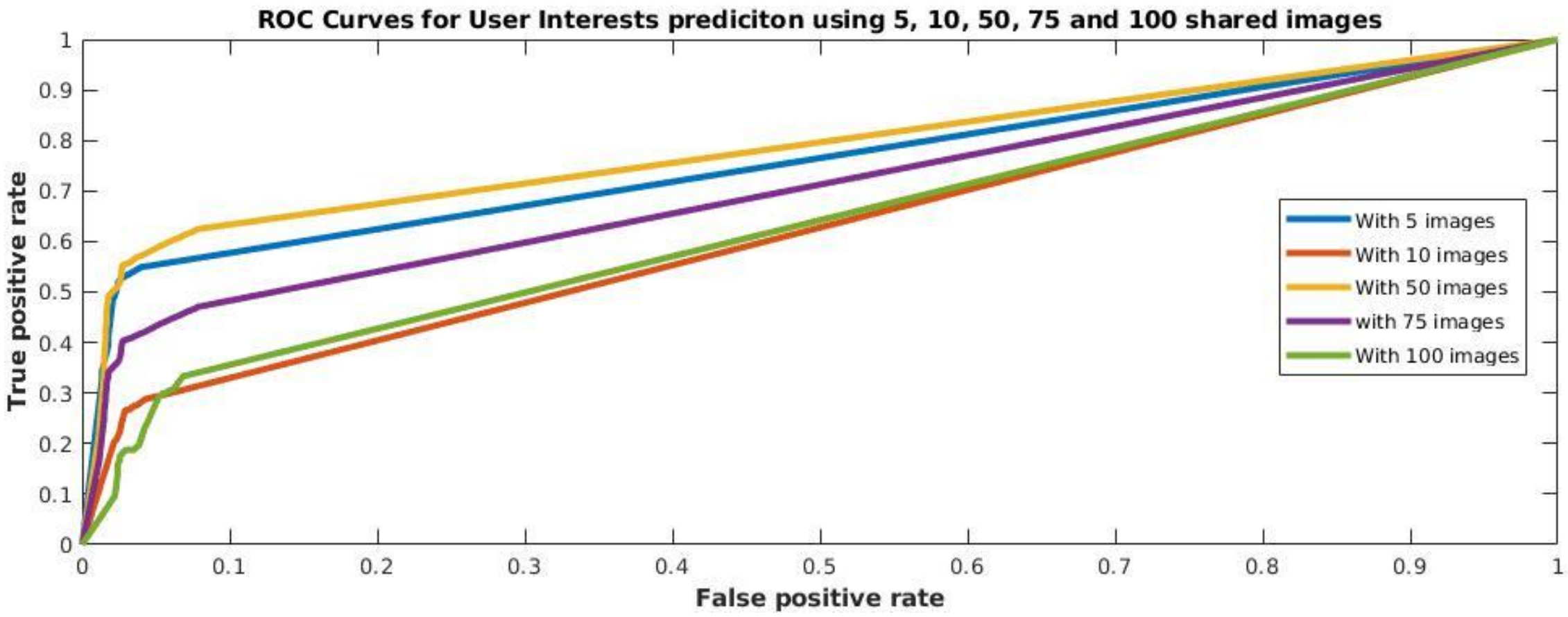

Fig. 33: Receiver Operating Characteristic (ROC) curve for users interests distribution

5 5 and 3 3 respectively for each convolutional layer. Furthermore, to achieve better performance, the complexity of convolutional neural networks is continually increasing with deeper architectures. The VGG'19 is much deeper than AlexNet with 19 layers including thirteen convolutional layers with filters size of 3 3 and 3 fully-connected layers. The use of very small filters sizes capture a set of fine deep visual features from the image input, decrease the parameters number and increase the filters number. The increase of filters number augment the depth of the input image and consequently the depth of the network which presents a critical component for good performance in the users interests classification task based on the complex scene image classification.

Finally, given that VGG'19 is based on the filters simplicity and the depth of the network, GoogleNet architecture is one of the first architecture that introduces the idea of executing the layers in parallel with the inception module based on the concatenation operation at different scale.

GoogleNet use 9 inception modules with a creative structuring of layers in order to improve performance and computationally efficiency. Hence, GoogleNet help us to get better classifica-tion accuracy by extracting information about the very fine grain details in the volume.

After having discussed the performance of our proposed framework as well as the efficiency of each used architecture for feature extraction module, we try to discuss the reasons why some images are misclassified over classes other than the self-assessed class. This misclassification is because of divers adversarial attacks in the from of delicate perturbations to each input user' image that conduct our framework to predict incorrect class compared to the self-assessed class. However, the users which have self-assessed classes like Relatinship, News, Wellness, Lifestyle and Hobies are more likely to have perturbations in their shared images that change the output class to any other target class.

Several examples of misclassified users are shown in figure 34. The labels above the selected images present the self-assessed classes for each group of users and the labels below the images are the target classes obtained by our framework.

In the first group on the right, we see a set of images shared by users, whose have News as self-assessed class. The classifica-tion of these images is based on the objects belong each one obtained by CNN architecture for objects recognition. These objects possess Fashion or Places or People as super class in our UIO ontology. This inference presents a perturbation to predict other target classes other than the self-assessed by the user.

In the second group in the middle, we illustrate some images shared by users whose have Relationship as self-assessed class. The classification of these images product Outdoors and People as target classes. This misclassification is caused by the fact that the objects belong these shared images have Outdoors or People as super class in our UIO ontology. In fact, the topic Relationship reflects the importance of making friendships with people in order to have a good time together.

In the third group on the left, we present divers images shared by users whose have Wellness as self-assessed topic. The fact that these shared images contains objects which have Outdoors or People, these users have Outdoors or People as target classes in our classification.

## VI. Conclusion

In this paper we examined the challenge of relating a set of visual features to topics of interests using a deep learning framework. A joint novel framework was established, named "DeepVisInterest", for predicting users' interests from social visual data applying mainly the CNN architectures for feature extractor and classification modules. We also contributed novel users' interests model for conceptualize and categorize the big five topics of interests into semantic representation using the ontology. Compared to previous works, the proposed framework is a hybrid one that takes advantages of divers CNN architectures to directly extract the most pertinent deep features from images for the user interests conceptualization and prediction. We systematically evaluated the proposed framework regarding our proposed social visual data database untitled SmartCityZenDB which contains over 24000 social image from 240 Facebook accounts. The results demonstrated the superiority of our approach over the literature in terms of both accuracy and the evolution of time.

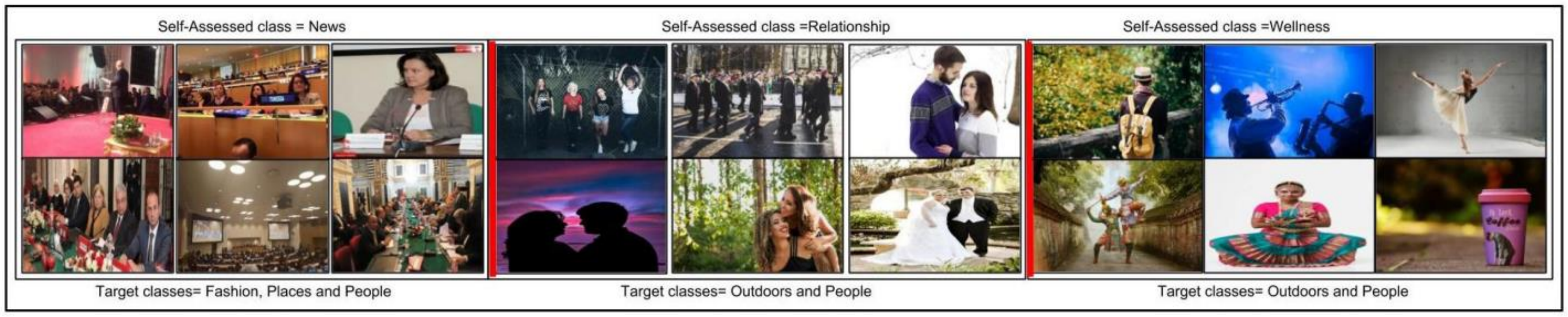

Fig. 34: Some adversarial examples of images: (a) Users with self-assessed class is News, (b) Users with self-assessed class is Relationship and (c) Users with self-assessed class is Wellness

## REFERÈNCIES